\documentclass[twocolumn,showpacs,aps,prd,fleqn]{revtex4}
\usepackage{fleqn}
\usepackage{dcolumn}
\usepackage{amsmath}
\usepackage{epsfig}
\usepackage{float}
\usepackage{relsize}
\usepackage{multirow}
\usepackage{graphicx}
\usepackage{mathrsfs}
\usepackage{epstopdf}
\input pubboard/babarsym

\newcommand{\BABARPubYear}    {14}
\newcommand{\BABARPubNumber}  {005}

\newcommand{\SLACPubNumber} {16123}

\newcommand{\proxy} {\textrm{proxy}}
\newcommand{\spare} {\textrm{spare}}
\newcommand{\soft} {\textrm{soft}}
\newcommand{\hard} {\textrm{hard}}

\def\figurebox#1#2#3{%
    \def\arg{#3}%
    \ifx\arg\empty
    {\hfill\vbox{\hsize#2\hrule\hbox to #2{\vrule\hfill\vbox to #1{\hsize#2\vfill}\vrule}\hrule}\hfill}%
    \else
    {\hfill\epsfbox{#3}\hfill}%
    \fi}

\begin{document}

\preprint{\babar-PUB-\BABARPubYear/\BABARPubNumber} 
\preprint{SLAC-PUB-\SLACPubNumber} 

\begin{flushleft}

\babar-PUB-\BABARPubYear/\BABARPubNumber\\
SLAC-PUB-\SLACPubNumber\\
\end{flushleft}

\title{
{\large \bf Bottomonium spectroscopy and radiative transitions involving the $\mathbf{\chi_{bJ}(1}\boldmath{P}\mathbf{,2}\boldmath{P)}$ states at \babar
 }
}

%
\author{J.~P.~Lees}
\author{V.~Poireau}
\author{V.~Tisserand}
\affiliation{Laboratoire d'Annecy-le-Vieux de Physique des Particules (LAPP), Universit\'e de Savoie, CNRS/IN2P3,  F-74941 Annecy-Le-Vieux, France}
\author{E.~Grauges}
\affiliation{Universitat de Barcelona, Facultat de Fisica, Departament ECM, E-08028 Barcelona, Spain }
\author{A.~Palano$^{ab}$ }
\affiliation{INFN Sezione di Bari$^{a}$; Dipartimento di Fisica, Universit\`a di Bari$^{b}$, I-70126 Bari, Italy }
\author{G.~Eigen}
\author{B.~Stugu}
\affiliation{University of Bergen, Institute of Physics, N-5007 Bergen, Norway }
\author{D.~N.~Brown}
\author{L.~T.~Kerth}
\author{Yu.~G.~Kolomensky}
\author{M.~J.~Lee}
\author{G.~Lynch}
\affiliation{Lawrence Berkeley National Laboratory and University of California, Berkeley, California 94720, USA }
% ...
\collaboration{The \babar\ Collaboration}
\noaffiliation


\begin{abstract}
We use $(121\pm1)$ million $\Upsilon(3S)$ and $(98\pm1)$ million $\Upsilon(2S)$ mesons recorded by the $\babar$ detector at the \pep2\ \epem\ collider at SLAC to perform a study of radiative transitions involving the $\chi_{b\mathrm{J}}(1P,2P)$ states in exclusive decays with $\mumu\gamma\gamma$ final states. We reconstruct twelve channels in four cascades using two complementary methods. In the first we identify both signal photon candidates in the Electromagnetic Calorimeter (EMC), employ a calorimeter timing-based technique to reduce backgrounds, and determine branching-ratio products and fine mass splittings. These results include the best observational significance yet for the $\chi_{b0}(2P)\to\gamma\Upsilon(2S)$ and $\chi_{b0}(1P)\to\gamma\Upsilon(1S)$ transitions. In the second method, we identify one photon candidate in the EMC and one which has converted into an $\epem$ pair due to interaction with detector material, and we measure absolute product branching fractions. This method is particularly useful for measuring $\Upsilon(3S)\to\gamma\chi_{b1,2}(1P)$ decays. Additionally, we provide the most up-to-date derived branching fractions, matrix elements and mass splittings for $\chi_b$ transitions in the bottomonium system. Using a new technique, we also measure the two lowest-order spin-dependent coefficients in the nonrelativistic QCD Hamiltonian. 
 \end{abstract}

\pacs{13.20.Gd, 14.40.Pq, 14.65.Fy}

\maketitle

\section{\boldmath Introduction}
\label{sec:BottomoniumIntroduction}
The strongly bound $\bbbar$ meson system -- bottomonium -- exhibits a rich positronium-like structure that is a laboratory for verifying perturbative and nonperturbative QCD calculations~\cite{ref:reviews}. Potential models and lattice calculations provide good descriptions of the mass structure and radiative transitions below the open-flavor threshold. Precision spectroscopy probes spin-dependent and relativistic effects. Quark-antiquark potential formulations have been successful at describing the bottomonium system phenomenologically~\cite{ref:reviews}. These potentials are generally perturbative in the short range in a Coulomb-like single-gluon exchange, and transition to a linear nonperturbative confinement term at larger inter-quark separation. The various observed bottomonium states span these two regions and so present a unique opportunity to probe these effective theories. 

Radiative transition amplitudes between the long-lived bottomonium states are described in potential models in a multipole expansion with leading-order electric and magnetic dipole -- $E1$ and $M1$ -- terms. The $E1$ transitions couple the $S$-wave $\Upsilon(nS)$ states produced in $\epem$ collisions to the spin-one $P$-wave $\chi_{b\mathrm{J}}(mP)$ states; suppressed $M1$ transitions are required to reach the spin-singlet states such as the ground-state $\eta_b(1S)$. 

The partial width for an $E1$ transition from initial state $i$ to final state $f$ is calculated in effective theories using~\cite{ref:skwarnicki}:
\begin{equation}\label{eqn:gammaE1}
  \Gamma_{i\to f}=\frac{4}{3} e_b^2\alpha C_{if}(2J_f+1)E_{\gamma}^3 | \langle n_f L_f | r | n_i L_i \rangle |^2,
\end{equation}
where $e_b$ is the charge of the $b$ quark, $\alpha$ is the fine-structure constant, $C_{if}$ is a statistical factor that depends on the initial- and final-state quantum numbers (equal to $1/9$ for transitions between $S$ and $P$ states), $E_{\gamma}$ is the photon energy in the rest frame of the decaying state, $r$ is the inter-quark separation, and $n$, $L$ and $J$ refer to the principal, orbital angular momentum and total angular momentum quantum numbers, respectively. Measurements of $E1$ transition rates directly probe potential-model calculations of the matrix elements and inform relativistic corrections.

Nonrelativistic QCD (NRQCD) calculations on the lattice have been used with success to describe the bottomonium mass spectrum in the nonperturbative regime~\cite{ref:gray, ref:monahan, ref:lepage, ref:r_lewis, ref:badalian}, including splittings in the spin-triplet $P$-wave states due to spin-orbit and tensor interactions. Experimental splitting results can be used as an independent check of the leading-order spin-dependent coefficients in the NRQCD Hamiltonian~\cite{ref:gray, ref:monahan}.  

\begin{figure}
  \begin{center}
    \includegraphics[width=0.40\textwidth]{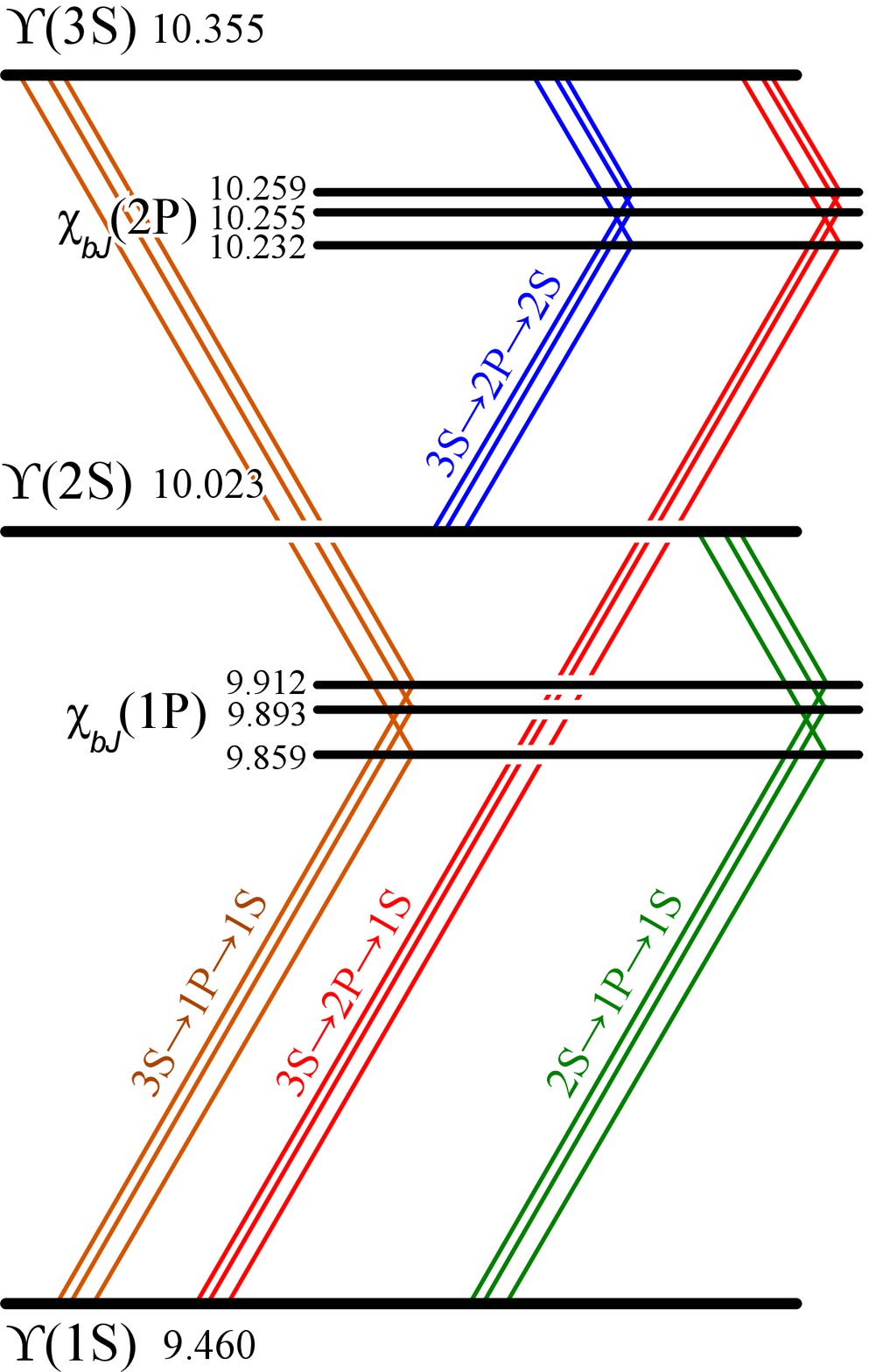}
    \vspace{-0.3cm}
    \caption{Schematic representation of the twelve $E1$ channels in the four radiative cascades measured in this analysis: $2S\to 1P\to 1S$, $3S\to 2P\to 2S$, $3S\to 1P\to 1S$, and $3S\to 2P\to 1S$. Each cascade terminates in the annihilation $\Upsilon(nS)\to\mumu$, not shown. The numbers give the masses \cite{ref:PDG} of the relevant bottomonium states in \gevcc. The first and second transitions in each cascade (except the $3S\to1P\to1S$) are referred to in the text as ``soft" and ``hard" transitions, respectively. Splittings in the photon spectra for these cascades are due to the mass splittings in the intermediate states $\chi_{bJ}$ with $J=0$, $1$ or $2$.  }
    \label{fig:Bottomonium}
    \vspace{-0.7cm}
\end{center}
\end{figure}

In the present analysis we measure radiative transition branching-ratio products and fine splittings in $E1$ transitions involving the $\chi_{bJ}(2P)$ and $\chi_{bJ}(1P)$ spin triplets, as displayed in Fig.~\ref{fig:Bottomonium}. We also provide relevant matrix elements and NRQCD coefficients for use in relativistic corrections and lattice calculations. These measurements are performed using two different strategies: in the first, we reconstruct the transition photons using only the {\babar\ }Electromagnetic Calorimeter (EMC); in the second, we consider a complementary set of such transitions in which one of the photons has converted into an $\epem$ pair within detector material.

Following an introduction of the analysis strategy in Sec.~\ref{sec:AnalysisIntroduction}, we describe relevant \babar\ detector and dataset details in Sec.~\ref{sec:babar}. The event reconstruction and selection, energy spectrum fitting, and the corresponding uncertainties for the calorimeter-based analysis are described in Sec.~\ref{sec:calo}. Sec.~\ref{sec:conv} similarly describes the photon-conversion-based analysis. Finally, we present results in Sec.~\ref{sec:results}, and a discussion and summary in Sec.~\ref{sec:summary}.

\section{\boldmath Analysis Overview}
\label{sec:AnalysisIntroduction}
Since the energy resolution of a calorimeter typically degrades with photon energy, the ${\sim20}\mevcc$ mass splittings of the $P$-wave bottomonium states are not resolvable for the ``hard'' ($\gtrsim200\mev$) $P\to S$ transitions but have been resolved successfully in ``soft'' ($\lesssim 200\mev$) $nS\to (n-1)P$ transitions by many experiments, including in high-statistics inclusive~\cite{ref:CUSB_2S_inclusive_softAndHard, ref:CrystalBall_2S_ inclusive_soft, ref:CUSB_3S_inclusive_soft, ref:CLEO_3S_inclusive_soft, ref:CLEO_2S_inclusive_splittings, ref:CLEO_3S_2S_inclusive_soft} and high-resolution converted~\cite{ref:CLEO_2S_conversion_soft, ref:ARGUS_2S_converted_soft} photon spectra. The hard transition rates are therefore less well-known, particularly for the $J=0$ states, which have large hadronic branching fractions. In particular, the individual $J=0$ hard transitions have been observed only by single experiments~\cite{ref:CUSB_3S_exclusive, ref:CUSB_3S_exclusive2, ref:CLEO_3S_2S_exclusive} and have yet to be confirmed by others. The $\Upsilon(3S)\rightarrow\gamma\chi_{b\mathrm{J}}(1P), \chi_{b\mathrm{J}}(1P)\to \gamma \Upsilon(1S)$ transitions are also experimentally difficult to measure because the soft and hard transition energies are nearly the same, and thus overlap. Previous measurements of $\BR(\Upsilon(3S)\rightarrow\gamma\chi_{b1,2}(1P))$ agree only marginally \cite{ref:CLEO_3S_2S_exclusive, ref:BABAR_conv}.

Two methods have been used to disentangle the $P$-wave spin states in the hard transitions: inclusive converted photon searches, used in a recent \babar\ analysis~\cite{ref:BABAR_conv}; and exclusive reconstruction of a two-photon cascade $S\to P\to S$ with dileptonic decay of the terminal $\Upsilon$~\cite{ref:CUSB_2S_exclusive, ref:CrystalBall_2S_exclusive, ref:CUSB_3S_exclusive, ref:CUSB_3S_exclusive2, ref:CLEO_3S_exclusive, ref:CLEO_3S_2S_exclusive}. In the first method, excellent energy resolution is achieved with a significant penalty in statistics. In the second method, the hard photon transitions are only indirectly measured, through their effect on the exclusive process. Here, we follow the latter strategy in an analysis of $E1$ transitions between bottomonium states below the open-flavor threshold in exclusive reconstruction of $\mumu\gamma\gamma$ final states. We use a large-statistics sample obtained by reconstructing the two photons in the cascade with the EMC to measure $\Upsilon(2S)\rightarrow\gamma\chi_{bJ}(1P)$, $\chi_{bJ}(1P)\rightarrow\gamma\Upsilon(1S)$ and $\Upsilon(3S)\rightarrow\gamma\chi_{bJ}(2P)$, $\chi_{bJ}(2P)\rightarrow\gamma\Upsilon(1S,2S)$ decays. We employ a background-reduction technique, new to {\babar\ }analyses, that utilizes EMC timing information. Furthermore, we reconstruct these same decay chains with one converted and one calorimeter-identified photon as a confirmation, and then extend this analysis to obtain a new measurement of $\Upsilon(3S)\rightarrow\gamma\chi_{bJ}(1P)$, $\chi_{bJ}(1P)\rightarrow\gamma\Upsilon(1S)$.

To simplify the notation, we hereinafter refer to the cascade $\Upsilon(2S)\to\gamma\chi_{bJ}(1P)$, $\chi_{bJ}(1P)\to\gamma\Upsilon(1S)$, $\Upsilon(1S)\to\mumu$ as $2S\to1P\to1S$ (and analogously for other cascades) where the muonic decay of the final state is implicit. Radiative photons are labeled based on the states that they connect: $\gamma_{2S\to 1P}$ and $\gamma_{1P\to 1S}$ for the example above. Unless noted otherwise all photon energies $E_{\gamma}$ are in the center-of-mass frame. The cascades measured in this analysis are shown in Fig.~\ref{fig:Bottomonium}. 

\section{\boldmath The \babar\ detector and dataset}
\label{sec:babar}
The \babar\ detector is described elsewhere~\cite{ref:NIM}, with the techniques associated with photon conversions described in Ref.~\cite{ref:BABAR_conv}. Only relevant details regarding the timing pipeline of the EMC are summarized here.

Energy deposited in one of the $6580$ CsI(Tl) crystals comprising the detector material of the EMC produces a light pulse that is detected by a photodiode mounted to the rear of the crystal. After amplification and digitization the pulse is copied onto a circular buffer which is read out upon arrival of a trigger signal. The energy-weighted mean of the waveform within a window encompassing the expected time of arrival of pulses is calculated and called the moment time. This moment time is compared to the event time -- the energy-weighted mean of all bins in the waveform above a threshold energy -- and the pulse is discarded if the difference between the two times is sufficiently large. For surviving waveforms, the moment time is bundled with the crystal energy and called a ``digi''. A collection of neighboring digis constitutes a ``cluster'' which can be associated with a neutral or charged particle candidate. The cluster time is a weighted mean of the digi times for all digis associated with a single cluster. 

Particle candidates are called ``in-time'' if they are part of an event that generates a trigger. The timing signature of an EMC cluster associated with an in-time event should be distinct from those for out-of-time events (primarily ``beam'' photons originating from interactions between the beam with stray gas or beam-related equipment, which are uncorrelated in time with events of physical interest). However, crystal-to-crystal differences (such as the shaping circuitry or crystal response properties) cause the quality of the EMC timing information to be low, and consequently it has been used only rarely to reject out-of-time backgrounds from non-physics sources. As a part of this analysis, we perform a calibration and correction of the EMC timing information. We present the results of an analysis of the performance of the corrected timing data in Sec.~\ref{sec:reco}.

The data analyzed include $(121\pm1)$ million $\Upsilon(3S)$ and $(98\pm1)$ million $\Upsilon(2S)$~\cite{ref:lumi} mesons produced by the PEP-II asymmetric-energy $e^+e^-$ collider, corresponding to integrated luminosities of $27.9\pm0.2\invfb$ and $13.6\pm0.1\invfb$, respectively. Large Monte Carlo (MC) datasets, including simulations of the signal and background processes, are used for determining efficiency ratios and studying photon line shape behavior. Event production and decays are simulated using \mbox{\tt Jetset \hspace{-0.5em}7.\hspace{-0.2em}4}\xspace~\cite{ref:JETSET} and \evtgen\ ~\cite{ref:EVTGEN}. We use theoretically predicted helicity amplitudes~\cite{ref:helicity} to model the angular distribution for each simulated signal channel, and we simulate the interactions of the final-state particles with the detector materials with \geant\ ~\cite{ref:GEANT}. 

\section{\boldmath Calorimeter-based Analysis}
\label{sec:calo}
\subsection{Event selection and reconstruction}
\label{sec:reco}
Candidate $mS\to (m-1)P\to nS$ cascades, with $m > n$ (that is, all cascades in Fig.~\ref{fig:Bottomonium} except $3S\to 1P\to 1S$), include $\mumu\gamma\gamma$ final states in data obtained at the $\Upsilon(mS)$ resonance, with both photons reconstructed using the EMC, and the four-particle invariant mass required to be within $300\mevcc$ of the nominal $\Upsilon(mS)$ mass. Photon candidates are required to have a minimum laboratory-frame energy of $30\mev$ and a lateral moment \cite{ref:LAT} less than $0.8$. A least-squares kinematic fit of the final-state particles under the signal cascade hypothesis is performed with the collision energy and location of the interaction point fixed. The dimuon mass is constrained to the $\Upsilon(nS)$ mass, and the $\mumu\gamma\gamma$ invariant mass is constrained to the $\Upsilon(mS)$ mass, both taken from the Particle Data Group~\cite{ref:PDG}. These constraints improve the soft photon resolution and allow better rejection of background from the decay $\Upsilon(mS)\to\piz\piz\Upsilon(nS)$, in which four final-state photons share the energy difference between the two $\Upsilon$ states, in contrast to the signal cascade which shares the same energy between only two photons. At this stage of reconstruction there are often many cascade candidates in each event; the $\chi^2$ probability from the kinematic fit is used to select the ``best'' candidate cascade in each event. The signal yields are obtained from a fit to the spectrum of the soft photon energy $E_{mS\to(m-1)P}$ in selected candidate cascades. 

Based on MC simulation of the soft photon spectrum, two significant background processes contribute: ``$\ppz$'' ($\Upsilon(mS)\to\ppz\Upsilon(nS)$; $\Upsilon(nS)\to\mumu$) and ``$\mu\mu(\gamma)$'' (continuum $\mumu$ production with initial- or final-state radiation or, rarely, $\Upsilon(mS)\to\mumu$ with QED bremsstrahlung photons). Regardless of the physics process, beam sources dominate the soft photon background.

To reject beam background we utilize the cluster timing information of the EMC. This is a novel technique not used in previous \babar\ analyses. We define the EMC cluster timing difference significance between two clusters $1$ and $2$ as $S_{1-2}\equiv |t_1-t_2|/\sqrt{\sigma_1^2+\sigma_2^2}$, where $t_i$ are the cluster times with associated timing uncertainties $\sigma_i$.  For background rejection we require $S_{\soft-\hard}<S_{\textrm{max}}$, where $S_{\textrm{max}}$ can be interpreted as the maximum allowable difference in standard deviations between the EMC timing of the soft and hard signal photon candidates. 

In conjunction with this analysis we have corrected several large nonuniformities in the EMC timing and calibrated the timing uncertainties; therefore characterization of the accuracy and precision of the timing information is required. To this end we use a proxy cascade mode which provides a precise and independent analog of the signal mode. Specifically, we reconstruct $\Upsilon(2S)\to \piz_{\proxy}\piz_{\spare}\Upsilon(1S)$, $\Upsilon(1S)\to\mumu$ cascades with final-state particles $\gamma^{\soft}_{\proxy}\gamma^{\hard}_{\proxy}\gamma^1_{\spare}\gamma^2_{\spare}\mu^{+}\mu^{-}$. The proxy $\piz_{\proxy}$ candidate is reconstructed from the proxy soft photon $\gamma^{\soft}_{\proxy}$ and the proxy hard photon $\gamma^{\hard}_{\proxy}$ candidates, which are required to pass the energy selections of the soft and hard signal photons. To remove mis-labeled cascades we reject events where the invariant mass of any combination of one proxy and one spare photon is in the $\piz$ mass range $100-155\mevcc$. A plot of the invariant mass of the $\piz_{\proxy}$ candidates now includes only two contributions: a peak at the nominal $\piz$ mass corresponding to in-time photon pairs and a continuous background corresponding to out-of-time photon pairs, almost exclusively the result of $\gamma^{\soft}_{\proxy}$ coming from beam background. We then measure the effect of the timing selection on in-time and out-of-time clusters by extracting the yields of these two contributions from fits over a range of $S_{\textrm{max}}$ values. We observe that the out-of-time rejection is nearly linear in $S_{\textrm{max}}$ and the functional form of the in-time efficiency is close to the ideal $\rm{erf}$$(S_{\textrm{max}} )$. With an EMC timing selection of $S_{\textrm{max}}=2.0$ we observe a signal efficiency of $0.92\pm0.02$ and background efficiency of $0.41\pm0.06$ in the proxy mode, and expect the same in the signal mode. 

\begin{figure}
  \begin{center}
    \includegraphics[width=0.47\textwidth]{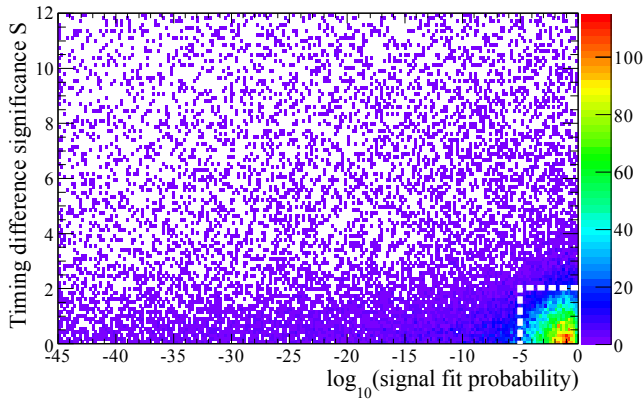}
    \vspace{-0.3cm}
    \caption{Scatter plot of reconstructed $2S\to 1P\to 1S$ events in the two selection variables $S_{\soft-\hard}$ and cascade kinematic fit probability for the calorimeter-based analysis. The cluster of in-time and high-probability events in the lower right corner is from the signal process, with a residue at lower probabilities due to tail events. The lack of structure in the scatter plot confirms the complementarity of these two selections. Events with $S_{\soft-\hard}>2.0$ or fit probability below $10^{-5}$ are excluded, as shown by the white dashed lines.}
    \label{fig:SvsP}
    \vspace{-0.7cm}
\end{center}
\end{figure}

To choose the ``best'' signal cascade candidate in an event we first require that the two photon energies fall within the windows $40-160\mev$ for $3S\to2P$, $160-280\mev$ for $2P\to2S$, $620-820\mev$ for $2P\to1S$, $40-240\mev$ for $2S\to1P$ or $300-480\mev$ for $1P\to1S$. Of these, only cascades with a timing difference significance between the two signal photon candidates below $2.0\sigma$  are retained: $S_{\soft-\hard}<S_{\textrm{max}}=2.0$ ($3.0\sigma$ for $3S\to2P\to1S$ to compensate for poorly-known timing uncertainties for higher photon energies). The best cascade candidate is further required to have a cascade fit probability in excess of $10^{-5}$, rejecting $90\%$ and $82\%$ of the passing $\piz\piz$ and $\mu\mu(\gamma)$ events according to reconstructions on MC simulations of those processes. The large majority of signal events lost in this selection have anomalously low-energy photon candidates which have deposited energy in the detector material that is not collected by the calorimeter. Excluding these events lowers the signal efficiency but improves our ability to disentangle the overlapping signal peaks during fitting. The highest-probability cascade candidate remaining in each event is then chosen. Fig.~\ref{fig:SvsP} demonstrates the selections on reconstructed $2S\to 1P\to 1S$ cascades.

\subsection{Fitting the photon energy spectra}
\label{sec:fitting}
We extract peak yield ratios and mean energy differences from the $E_{mS\to (m-1)P}$ spectra using unbinned maximum likelihood fits with three incoherent overlapping signal components corresponding to the $J=0$, $1$ and $2$ decay channels and a smooth incoherent background. Simulated signal, and $\mu\mu(\gamma)$ and $\piz\piz$ background MC collections are subjected to the same reconstruction and selection criteria as the data, scaled to expected cross sections, and combined to constitute the ``MC ensemble'' which is representative of the expected relevant data. Qualitative agreement between the MC ensemble and data spectra is good, although the MC line shapes deviate from the data line shapes enough that fit solutions to the individual MC lines cannot be imposed on the corresponding fits to the data. 

A fit to an individual peak from a signal MC collection requires the flexibility of a double-sided Crystal Ball~\cite{ref:CB} fitting function. This function has a Gaussian core of width $\sigma$ and mean $\mu$ which transitions at points $\alpha_1$ and $\alpha_2$ to power-law tails with powers $n_1$ and $n_2$ on the low- and high-energy sides, respectively, with the requirement that the function and its first derivative are continuous at the transition points. The background spectrum of the MC ensemble is described well by the sum of a decaying exponential component with power $\lambda$ and a linear component with slope $a_1$. The simplest approach to fitting the spectrum is to float both background parameters $\lambda$ and $a_1$ and float Gaussian means for the three signal peaks $\mu_0$, $\mu_1$ and $\mu_2$ separately while sharing the floated signal shape parameters $\sigma$, $\alpha_1$, $\alpha_2$, $n_1$ and $n_2$ between all three signal peaks. This approach assumes that the line shape does not vary in the limited photon energy range of this spectrum. However, fits of this nature on the MC ensemble spectrum perform poorly, indicating that line shape variation cannot be ignored. Conversely, fits with all twenty signal and background parameters floating independently perform equally poorly; in particular, the $J=0$ peak tends to converge to a width well above or below the detector resolution. A more refined fitting strategy is required. 

\begin{figure*}
  \centering
    \includegraphics[height=5.5in]{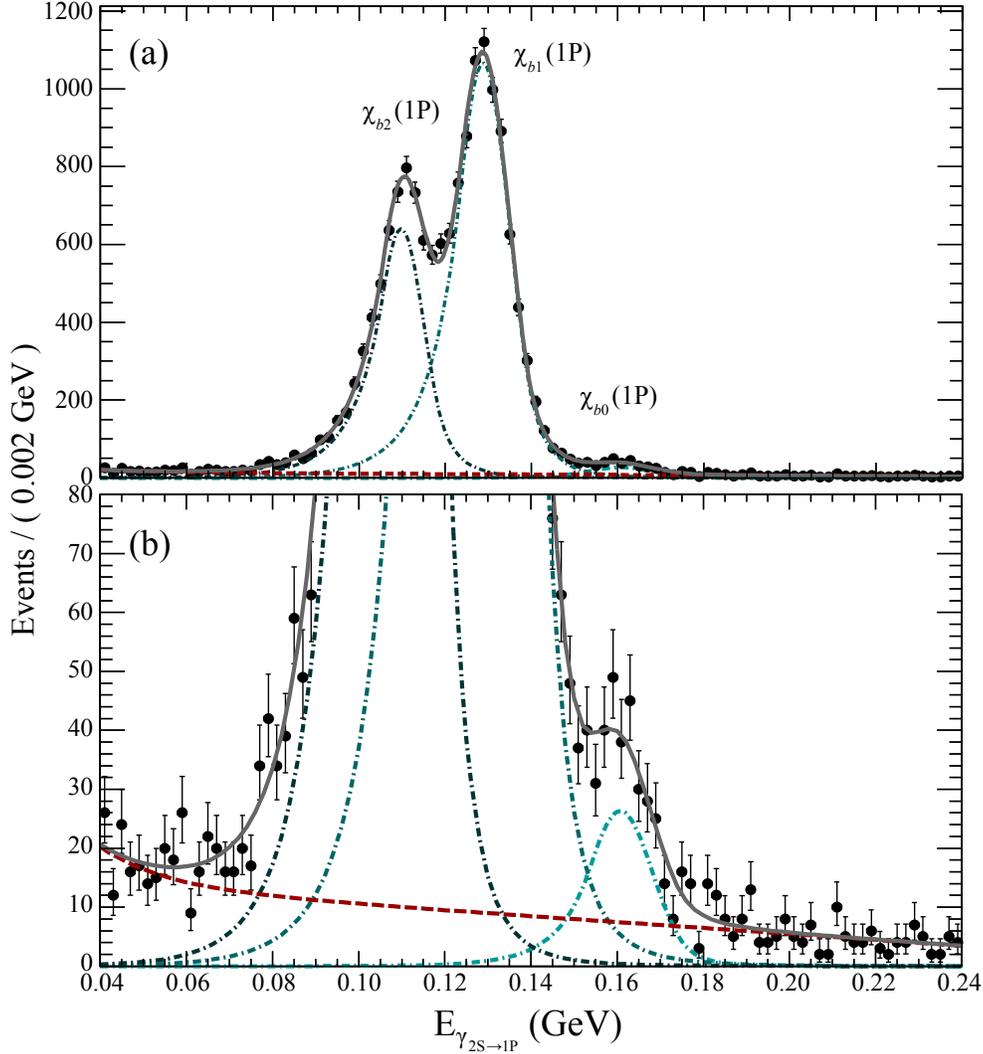}
    \caption{(a) Fit to the soft photon energy $E_{2S\to 1P}$ in the $2S\to 1P\to 1S$ cascade with individual signal (dot-dashed) and background (dash) components for the calorimeter-based analysis. The targeted $\chi_{b0}(1P)$ signal corresponds to the small bump on the right; the integral ratio and mean offset of the fit to this peak compared to the $J=1$ peak (center), are defined as $f_0$ and $\delta_0$, respectively. Similarly, $f_2$ and $\delta_2$ are the integral ratio and mean offset for the fit to the $J=2$ peak (left), also compared to the $J=1$ peak. (b) A zoomed-in view of the $\chi_{b0}(1P)$ peak on the same energy scale.}
    \label{fig:Y2S_fit}  
\end{figure*}

\begin{figure*}
  \centering
    \includegraphics[height=5.5in]{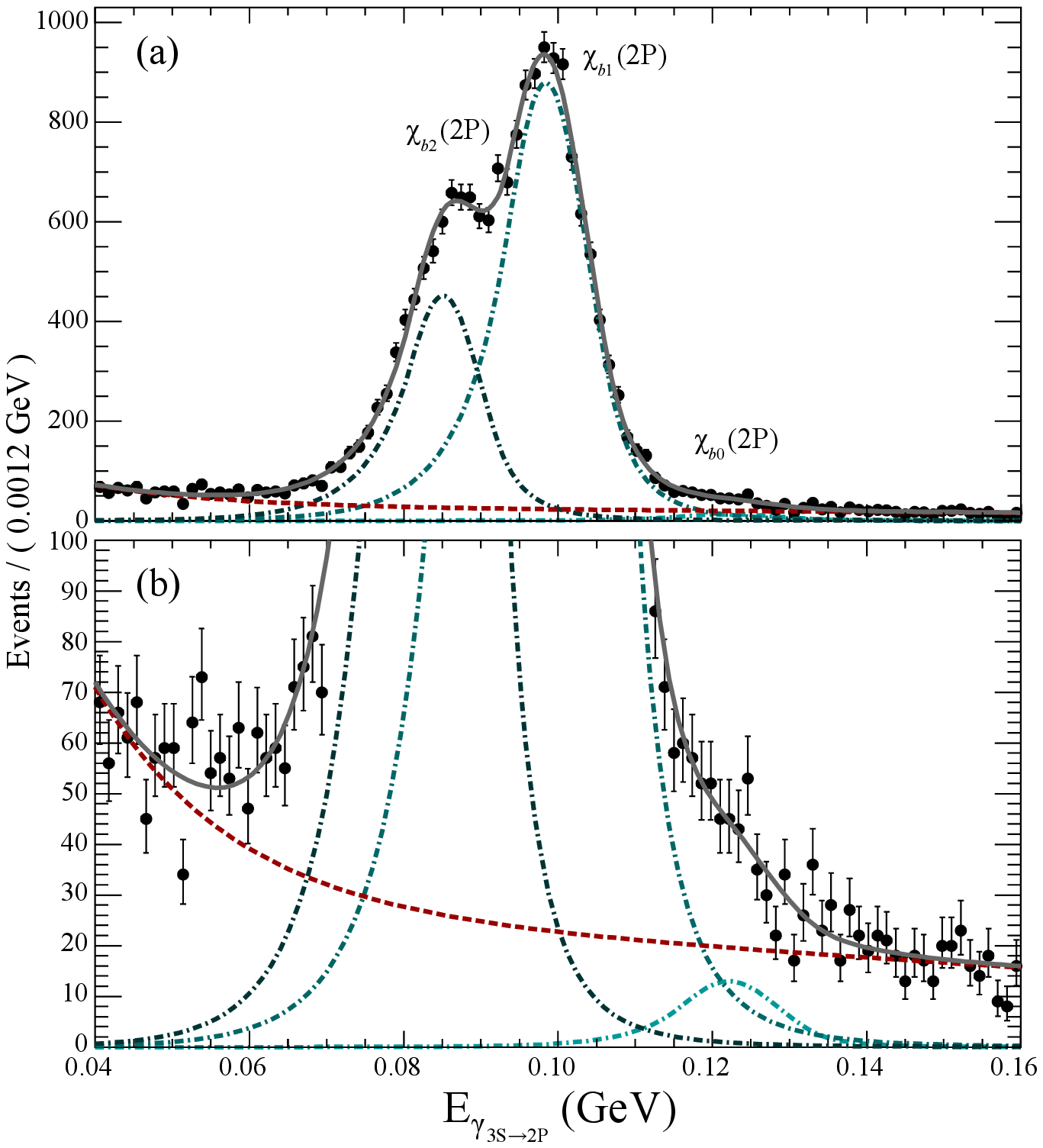}
    \caption{(a) Fit to the soft photon energy $E_{3S\to 2P}$ in the $3S\to 2P\to 2S$ cascade with individual signal (dot-dashed) and background (dash) components for the calorimeter-based analysis. (b) A zoomed-in view of the $\chi_{b0}(2P)$ peak on the same energy scale. Discussion of the significance of the $J=0$ peak is contained in Section~\ref{sec:systematics}.}
    \label{fig:Y3S_2S_fit}  
\end{figure*}

\begin{figure*}
  \centering
    \includegraphics[height=5.5in]{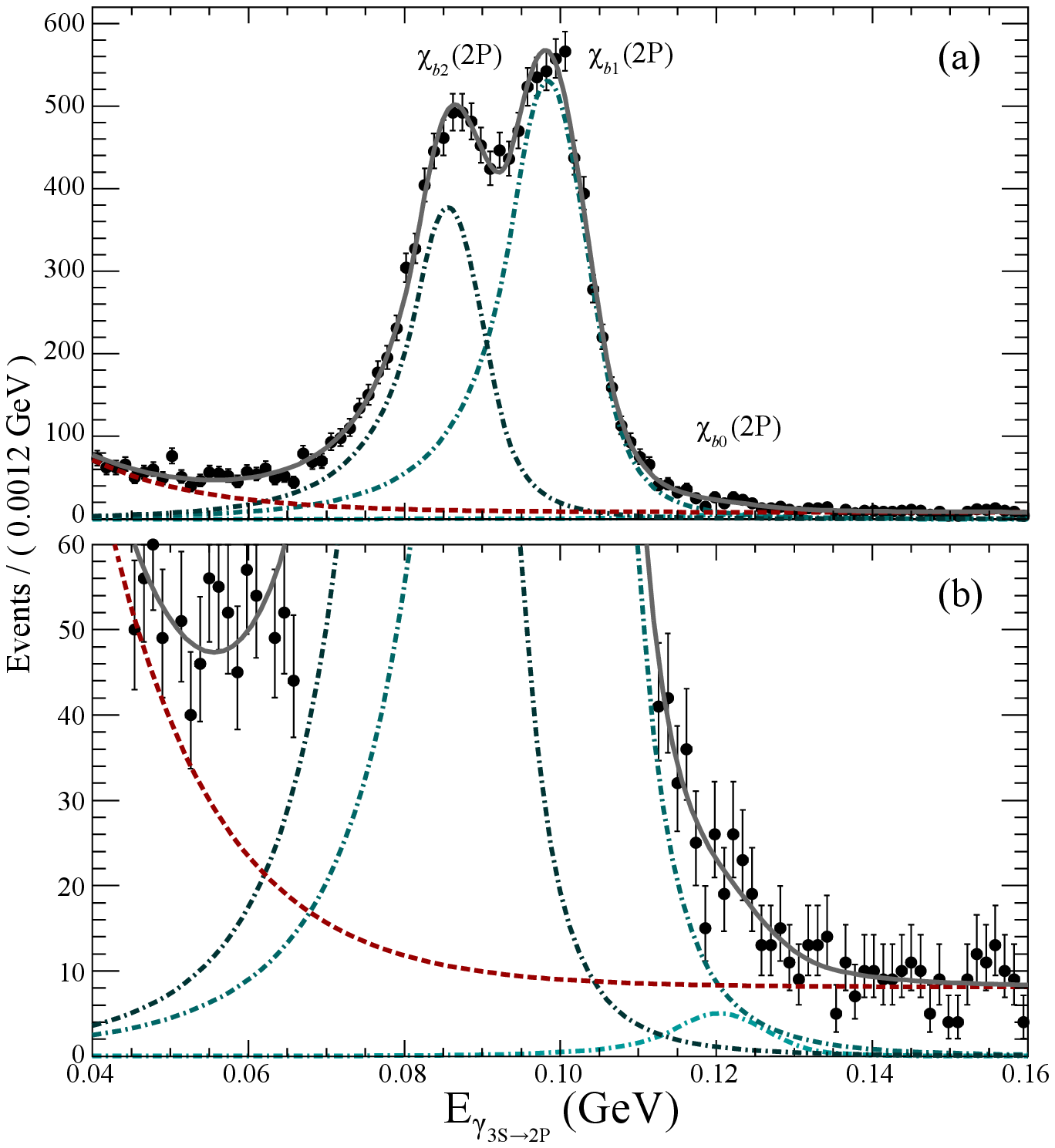}
    \caption{(a) Fit to the soft photon energy $E_{3S\to 2P}$ in the $3S\to 2P\to 1S$ cascade with individual signal (signal corresponds to the small bump) and background (dash) components for the calorimeter-based analysis. (b) A zoomed-in view of the $\chi_{b0}(2P)$ peak on the same energy scale. Discussion of the significance of the $J=0$ peak is contained in Section~\ref{sec:systematics}.}
    \label{fig:Y3S_1S_fit}  
\end{figure*}

To obtain stable fits to the data spectrum, we allow the parameters $\sigma$, $\alpha_1$ and $\alpha_2$ of the dominant $J=1$ peak to float, and we fix the corresponding $J=0$ and $J=2$ parameters with a linear extrapolation from the $J=1$ values using slopes derived from fits to the MC signal spectra. The fit is insensitive to the power of the tails, so $n_1$ and $n_2$ are fixed to solutions from fits to signal MC. The background parameters $\lambda$ and $a_1$ are fixed to MC solutions but the ratio of background contributions floats, as does the absolute background yield $N_{bkg}$. 
The signal component functions are expressed in terms of the desired observables: signal yield ratios $f_J=N_J/N_1$ and peak mean offsets $\delta_J=\mu_J-\mu_1$, which both float in the fit, as well as the $J=1$ yield, $N_1$, and mean, $\mu_1$. Figures~\ref{fig:Y2S_fit},~\ref{fig:Y3S_2S_fit} and~\ref{fig:Y3S_1S_fit} show the results of the fits to the three data spectra.

\subsection{Systematic studies}
\label{sec:systematics}

We measure branching-ratio products for cascades involving $\chi_{b\mathrm{J}}$ normalized to the $\chi_{b1}$ channel and denote them $\mathscr{F}^{J/1}$. In this way we avoid the systematic uncertainty associated with estimating absolute reconstruction and selection efficiencies, which cancel in the ratio. In terms of measured values, the exclusive branching ratio is given by
\begin{equation} \label{eqn:brratio}
  \begin{split}
\mathscr{F}^{J/1}_{mS\to P(J) \to nS}  &= \frac{\BR(mS\to P(J))\BR(P(J)\to nS)}{\BR(mS\to P(1))\BR(P(1)\to nS)} \\
                                       &= f_J\frac{\epsilon_1}{\epsilon_J}
  \end{split}
\end{equation}
where $\epsilon_{J}$ is the signal efficiency of the $J$ channel and the measured yield ratio $f_J$ has systematic corrections applied. The branching fraction of the terminal $\Upsilon(nS)\to\mumu$ decay appears in both the numerator and denominator and thus cancels. We also measure the mass splittings $\Delta M_{J-1}$, which are simply equal to the measured peak energy differences $\pm\delta_J$ with systematic corrections. In this way we avoid the systematic uncertainties associated with determining the absolute photon energies. Systematic effects and uncertainties on the yield ratio $f_J$, line energy differences $\delta_{\mathrm{J}}$, and efficiency ratio $\epsilon_{1} / \epsilon_{J}$ are discussed below.

Constraining the line shape parameters to fixed linear slopes introduces unknown systematic biases in the extracted yield and mean values, resulting in systematic uncertainties. To measure these uncertainties, a collection of $50,000$ model spectra is generated that violates these assumptions; each spectrum is fit with the same fitting procedure as for the data spectrum. The behavior of the fits to these generic model spectra constrains the uncertainty of the fit to the data spectrum. 
The functions used to generate the model spectra are taken from the fit to the data spectrum with all parameters varying in a flat distribution within $\pm3\sigma$ of their fitted values, with these exceptions: the tail power parameters are varied in the range $5.0-100.0$ and the parameter slopes, taken from MC, are varied within $\pm5\sigma$ of their nominal values. The three peaks are decoupled to violate the single-slope fitting constraint. 

The fitting procedure fails to converge for some of the generated spectra. These spectra are evidently not sufficiently similar to the data spectrum and can be discarded without biasing the set of generated spectra. We further purify the model spectrum collection by rejecting models with fitted parameters outside $\pm3\sigma$ of the data fit solution. For the successful fits we define the pull for a parameter $X$ ($N$ or $\mu$) as $(X_{\textrm{generated}}-X_{\textrm{fit}})/\sigma_X$, where $\sigma_X$ is the parameter uncertainty in the fit. We fit a Gaussian function to each pull distribution for the surviving model fits and observe a modest shift in central value and increase in width (see Table~\ref{tab:pulls}). We use this shift to correct the data fit parameter values, and scale the parameter uncertainties by the width of the pull distribution. In this way we have used the model spectra to measure systematic uncertainties and biases associated with the fitting procedure, and used these to correct the data fit results. 

Two considerations arise in interpreting these scaled uncertainties. First, statistical and systematic sources of uncertainty are admixed and cannot be disentangled. Second, all of the uncertainties are necessarily over-estimated. However, the overestimation of uncertainty is smaller than the difference between scaled and unscaled uncertainty, which is itself much less than one standard deviation (see Table~\ref{tab:pulls}). The model spectrum selection procedure guarantees further that the overestimation is limited, and in fact further tightening the selections does not decrease the width of the pull distributions, indicating convergence. We conclude that the overestimation of uncertainties is negligible. 

The absolute signal efficiencies are a combination of unknown reconstruction and selection efficiencies with attendant systematic uncertainties which cancel in the ratio. The efficiency ratio in Eq.~(\ref{eqn:brratio}) deviates from unity due to spin-dependent angular distributions in a detector with non-isotropic acceptance. The signal MC collections simulate the model-independent angular distributions of the decay products in the signal cascades for the three $1P$ spin states~\cite{ref:helicity} as well as the detector response. Uncertainties in the ratio come from two sources: MC sample size and the effect of the fit probability selection on the ratio. We measure the ratio in signal MC and add in quadrature the standard deviation of the ratio taken over a variety of fit probability selections as an estimation of the efficiency ratio uncertainty (see Table~\ref{tab:efficiencies}).

{
  \begin{table}
    \caption{MC efficiency ratios for the calorimeter-based analysis.}
    \begin{center}
      \renewcommand{\tabcolsep}{4pt}
      \renewcommand{\arraystretch}{1.2}
      \begin{tabular}{ l l l }
          Cascade                 & $J$ & $\epsilon_{1} / \epsilon_{J}$    \\ \hline \hline \\ [-2.0ex]
          $2S\to1P\to1S$  & $0$ & $1.062\pm0.009$         \\
				      & $2$ & $1.013\pm0.004$        \\
          $3S\to2P\to2S$  & $0$ & $1.059\pm0.005$         \\
				      & $2$ & $0.988\pm0.003$        \\
          $3S\to2P\to1S$  & $0$  & $1.059\pm0.009$         \\
				      & $2$ & $1.027\pm0.009$        \\
        \end{tabular}
      \label{tab:efficiencies}
  \end{center}
\end{table}
}
{
  \begin{table*}
    \caption{Results of systematic studies on $50,000$ model spectra for each of the three signal channels in the calorimeter-based analysis, with parameter values in units of $10^{-2}$ for $f_J$ and $\mev$ for $\delta_J$. For the $3S\to2P\to2S$ and $3S\to2P\to1S$ analyses negative- and zero-yield models for the $J=0$ peaks are inconsistent with the fit results with a significance of $5.1\sigma$ and $2.1\sigma$, respectively. The efficiency ratio corrections have not been applied.}
    \begin{center}
      \renewcommand{\tabcolsep}{4pt}
      \renewcommand{\arraystretch}{1.2}
      \begin{tabular}{ l c D{,}{\,\pm\,}{-1} c c D{,}{\,\pm\,}{4.4}}
          Cascade           
          & Parameter  
          & \multicolumn{1}{c}{Fit value}     
          & Pull shift $(\sigma)$ 
          & Pull width $(\sigma)$ 
          & \multicolumn{1}{c}{Corrected}      \\ \hline \hline \\ [-2.0ex]
          $2S\to1P\to1S$ & $f_0$      & 2.83,0.31   & $+0.82$               & $1.12$                & 3.09,0.35   \\
                         & $f_2$      & 54.8,1.3    & $+0.078$              & $1.15$                & 54.9,1.5    \\
                         & $\delta_0$ & 32.00,0.91   & $+0.54$               & $1.03$                & 32.5,0.93   \\
                         & $\delta_2$ & -19.00,0.22 & $-0.036$              & $1.06$                & -19.01,0.24 \\ \hline
          $3S\to2P\to2S$ & $f_0$      & 1.66,0.39   & $+1.5$                & $1.13$                & 2.25,0.44   \\
                         & $f_2$      & 47.7,1.3    & $-0.47$               & $1.29$                & 47.0,1.7    \\
                         & $\delta_0$ & 22.60,0.20   & $+0.54$               & $1.03$                & 23.7,2.1    \\
                         & $\delta_2$ & -13.30,0.22 & $-0.036$              & $1.06$                & -13.3,0.24  \\ \hline
          $3S\to2P\to1S$ & $f_0$      & 1.05,0.52   & $+1.1$                & $1.44$                & 1.62,0.75   \\
                         & $f_2$      & 66.3,2.3    & $-0.63$               & $1.27$                & 64.9,2.9    \\
                         & $\delta_0$ & 21.60,0.30   & $+0.80$               & $1.13$                & 24.0,3.4    \\
                         & $\delta_2$ & -12.80,0.22 & $+0.032$              & $1.03$                & -12.79,0.23 \\
        \end{tabular}
      \label{tab:pulls}
  \end{center}
\end{table*}
}
\section{\boldmath Converted Photon Analysis}
\label{sec:conv}
\subsection{Event Selection and Reconstruction}
In the conversion-based analysis the $\mumu\gamma\gamma$ final state is reconstructed by requiring one of the photons to be identified in the EMC and the other to be reconstructed after converting into an \epem pair in detector material. Although it shares the same underlying physics as the calorimeter-based analysis, the presence of a displaced vertex and lack of calorimeter timing information necessitate some differences in approach.

We reconstruct the $\Upsilon(1S,2S)$ final states with two opposite-sign muons within 100 $\mevcc$ of the relevant $\Upsilon(nS)$ mass, satisfying a vertex probability $\chi^{2}$ of greater than 0.0001. The $\chi_{bJ}(mP)$ candidates are formed by constraining the $\Upsilon(nS)$ to its nominal mass \cite{ref:PDG} and adding a converted photon (as described in detail in Ref.~\cite{ref:BABAR_conv}). The initial $\Upsilon(2S,3S)$ candidate is reconstructed by combining a calorimeter-identified photon candidate with the $\chi_{bJ}(mP)$ candidate. This photon is required to have a minimum laboratory-frame energy of 30\mev and lateral moment less than 0.8. The center-of-mass energy of the calorimeter photon is required to be in the range $300<E_{\gamma}<550$\mev for the $3S\rightarrow 1P\rightarrow 1S$ decay chain, while for the other transitions it must be within $E_{\gamma(\textrm{low})}-40 < E_{\gamma} < E_{\gamma(\textrm{high})}+40$\mev, where $E_{\gamma(\textrm{low})}$ and $E_{\gamma(\textrm{high})}$ represent the lowest and highest energy transition for the intermediate $\chi_{bJ}(mP)$ triplet in question.

Because this reconstruction lacks a sufficient second EMC timing measurement to take advantage of the timing-based selection described in Sec.\ref{sec:reco}, $\pi^{0}\pi^{0}$ and $\mumu(\gamma)$ backgrounds, which are also the dominant background sources for this analysis technique, are reduced via more conventional means. The number of charged-particle tracks in the event, as identified by the {\babar\ }drift chamber and silicon vertex tracker~\cite{ref:NIM}, is required to be equal to four, incidentally removing all events used in the calorimeter-based analysis and resulting in mutually exclusive datasets. This selection also makes this dataset independent from the previous inclusive converted photon analysis~\cite{ref:BABAR_conv}, with the only commonality being shared uncertainties on the luminosity measurement and the conversion efficiency (described in Sec.~\ref{sec:conv_systUnc}). Events with an initial $\Upsilon(nS)$ mass, $M_{\Upsilon(nS)}$, in the range $10.285<M_{\Upsilon(3S)}<10.395$\gevcc or $M_{\Upsilon(2S)(\textrm{PDG})}\pm40$\mevcc are retained. A requirement that the ratio of the second and zeroth Fox-Wolfram moments \cite{ref:foxwolfram} of each event, $R_{2}$, be less than 0.95 is also applied. These selection criteria were determined by maximizing the ratio of expected $3S\rightarrow 1P\rightarrow 1S$ signal events to the square root of the sum of the expected number of signal and background events, as determined by MC simulation.

We calculate the signal efficiency by counting the number of MC signal events remaining after reconstruction and the application of event selection criteria. The efficiency is highly dependent upon the conversion photon energy (as seen in Ref.~\cite{ref:BABAR_conv}), and ranges from $0.1\%$ at $E_{\gamma}{\sim}200$\mev to $0.8\%$ at $E_{\gamma}{\sim}800$\mev. This drop in efficiency at lower energies makes a measurement of the soft photon transitions impractical with converted photons, which is why this analysis is restricted to conversion of the hard photon. Once the full $\Upsilon(nS)$ reconstruction is considered, the overall efficiency ranges from $0.07-0.90\%$ depending on the decay chain. While the efficiency for the reconstruction with a converted photon is low, this technique leads to a large improvement in energy resolution from approximately $15\mev$ to $2.5\mev$. This is necessary in order to disentangle the transition energy of the hard photon from the overlapping signals for the $3S\rightarrow 1P\rightarrow 1S$ transitions. However, despite this improvement in energy resolution, the mass splittings are not measured with this technique because of line shape complications described in the following section.

\subsection{Fitting}
We use an unbinned maximum likelihood fit to the hard converted photon spectrum to extract the total number of events for each signal cascade. In the case of $3S\to 1P\to 1S$ transitions, the first and second transitions overlap in energy and either photon may be reconstructed as the converted one. Therefore both components are fit simultaneously. Because we analyze the photon energy in the center-of-mass frame of the initial $\Upsilon(nS)$ system, the photon spectra from subsequent boosted decays (\emph{e.g.} $\chi_{bJ}(mP)\rightarrow\gamma\Upsilon(nS)$) are affected by Doppler broadening due to the motion of the parent state in the center-of-mass frame. Due to this effect, variation of efficiency over the photon angular distribution, and a rapidly changing converted photon reconstruction efficiency, the signal line shapes are most effectively modeled using a kernel estimation of the high statistics MC samples. This is most relevant for the $3S\rightarrow 1P\rightarrow 1S$ transitions, which are the focus of this part of the analysis, and for which the signal line shape for the $1P\to 1S$ transition in $3S\to 1P\to 1S$ is so significantly Doppler-broadened that its shape can be qualitatively described by the convolution of a step-function with a Crystal Ball function. Alternative parameterizations using variations of the Crystal Ball function as described in Sec.~\ref{sec:fitting}, give a good description of the other transition data, but are reserved for evaluation of systematic uncertainties in this analysis. 

The MC simulation indicates the presence of a smooth \ppz and $\mumu(\gamma)$ background below the signal peaks. This primarily affects the $3S\rightarrow 2P\rightarrow (1S,2S)$ cascades, but is also present for $2S\rightarrow 1P\rightarrow 1S$. The background is modeled by a Gaussian with a large width and a mean above the highest transition energy for each triplet. For $3S\rightarrow 1P\rightarrow 1S$, both photons are hard and therefore the background is expected to be much smaller, and to have a flatter distribution. It is modeled with a linear function.

To allow for potential line shape differences between the simulation and data, energy scale and resolution effects are considered both by allowing the individual signal peak positions to shift and by applying a variable Gaussian smearing to the line shape. These effects are determined from the fit to the higher-statistics $J=1$ and $J=2$ peaks in the $3S\rightarrow 2P\rightarrow 1S,2S$ and $2S\rightarrow 1P\rightarrow 1S$ analysis energy regions, and the yield-weighted average for the energy scale shift and resolution smearing are applied to the $3S\rightarrow 1P\rightarrow 1S$ fit. The applied peak shift correction is $-0.1$\mev, with maximal values ranging from $-1.5$ to $0.9$\mev, and the required energy resolution smearing is less than $0.2$\mev. These energy scale values are consistent with those found in the previous, higher-statistics, {\babar\ }inclusive converted photon analysis \cite{ref:BABAR_conv}, and the resolution smearing is small compared to the predicted resolution, which is of the order of a few~\mev.

\begin{figure*}
  \centering
    \includegraphics[height=2.75in]{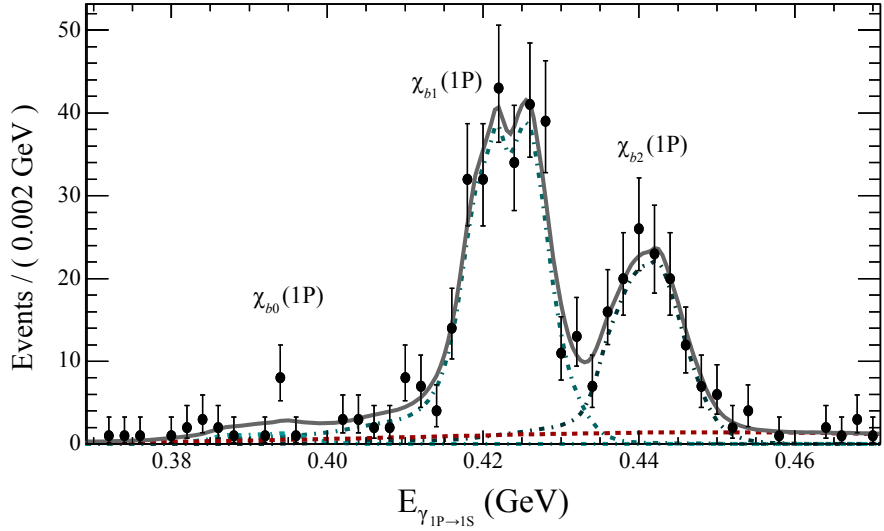}
    \caption{Fit to the photon energy $E_{1P\to 1S}$ in the $2S\to 1P\to 1S$ cascade for the conversions analysis. The data are represented by points, the total fit by a solid curve, the background component with a dashed curve, and the individual signal components of the fit by peaking dot-dashed curves in progressively darker shades for $J=0,1,2$. Clear evidence is seen for the $J=2$ and $J=1$ signals.}
    \label{fig:conv2s}
\end{figure*}

\begin{figure*}
  \centering
    \includegraphics[height=2.75in]{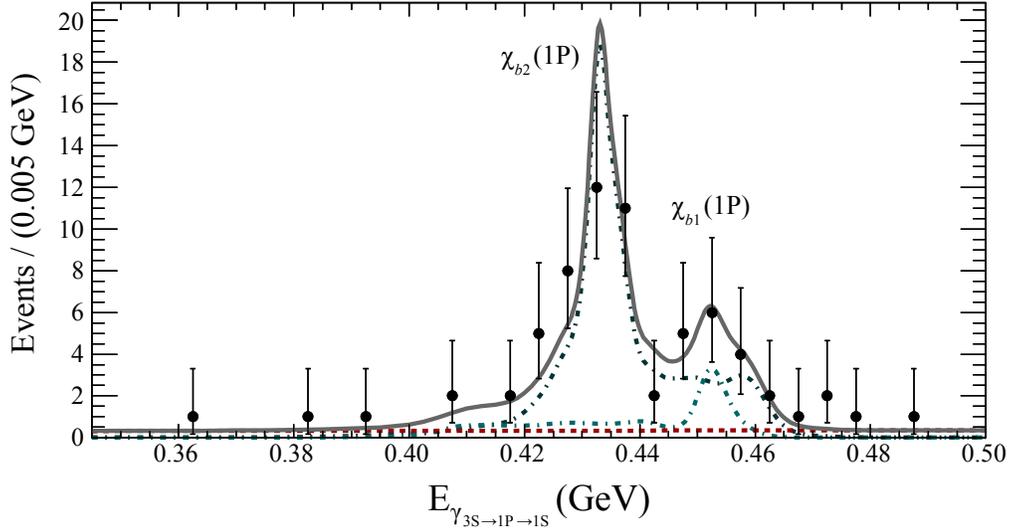}
    \caption{Fit to the photon energy $E_{3S\to 1S}$ in the  $3S\to 1P\to 1S$ cascade for the conversion analysis. The data are represented by points, the total fit by a solid curve, the background component with a dotted curve, and the individual signal components of the fit by peaking dot-dashed curves for $J=2$ (darker shade) and $J=1$ (lighter shade). There is clear evidence for the $J=2$ transition, but the $J=1$ signal is not statistically significant.}
    \label{fig:conv3s}
\end{figure*}

\begin{figure*}
  \centering
    \includegraphics[height=2.75in]{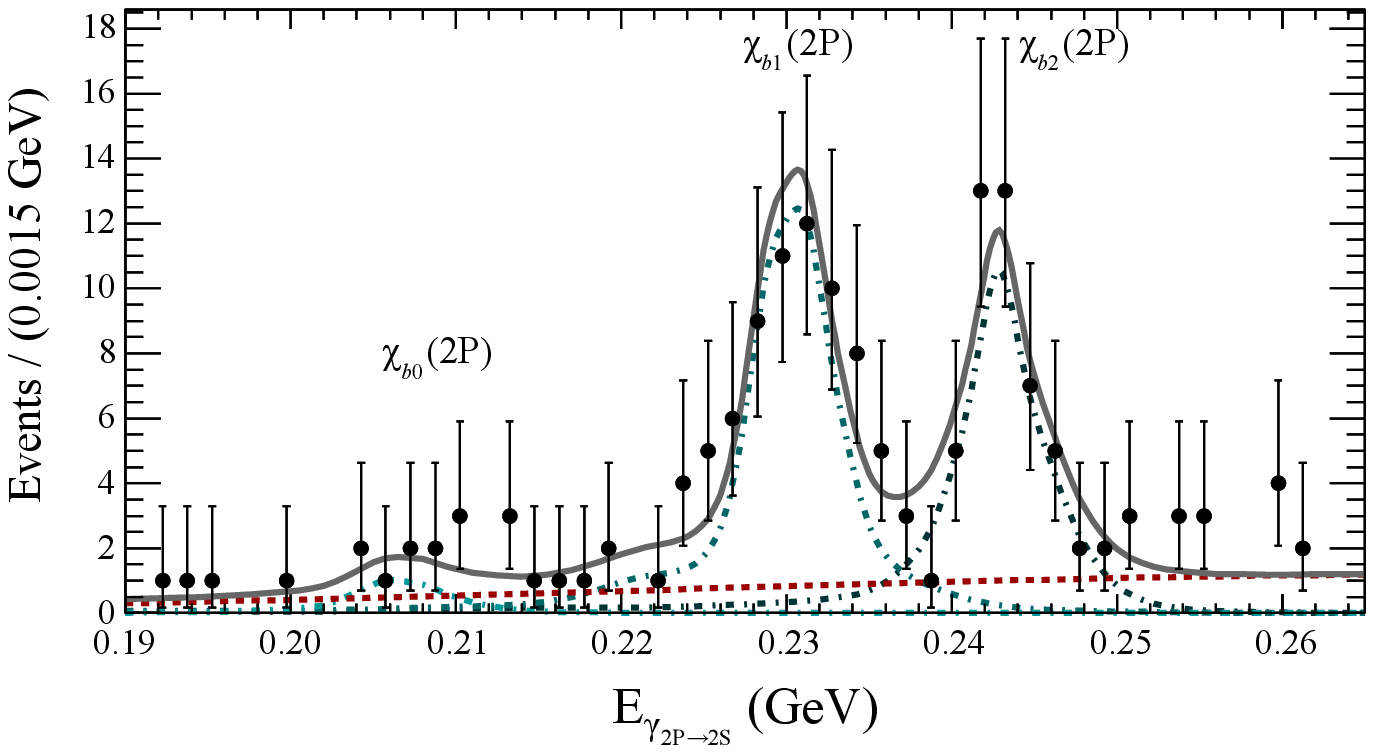}
    \caption{Fit to the photon energy $E_{2P\to 2S}$ in the $3S\to 2P\to 2S$ cascade for the conversion analysis. The data are represented by points, the total fit by a solid curve, the background component with a dotted curve, and the individual signal components of the fit by peaking dot-dashed curves in progressively darker shades for $J=0,1,2$. Clear evidence for the $J=2$ and $J=1$ signals are seen.}
    \label{fig:convlo}
\end{figure*}

\begin{figure*}
  \centering
    \includegraphics[height=2.75in]{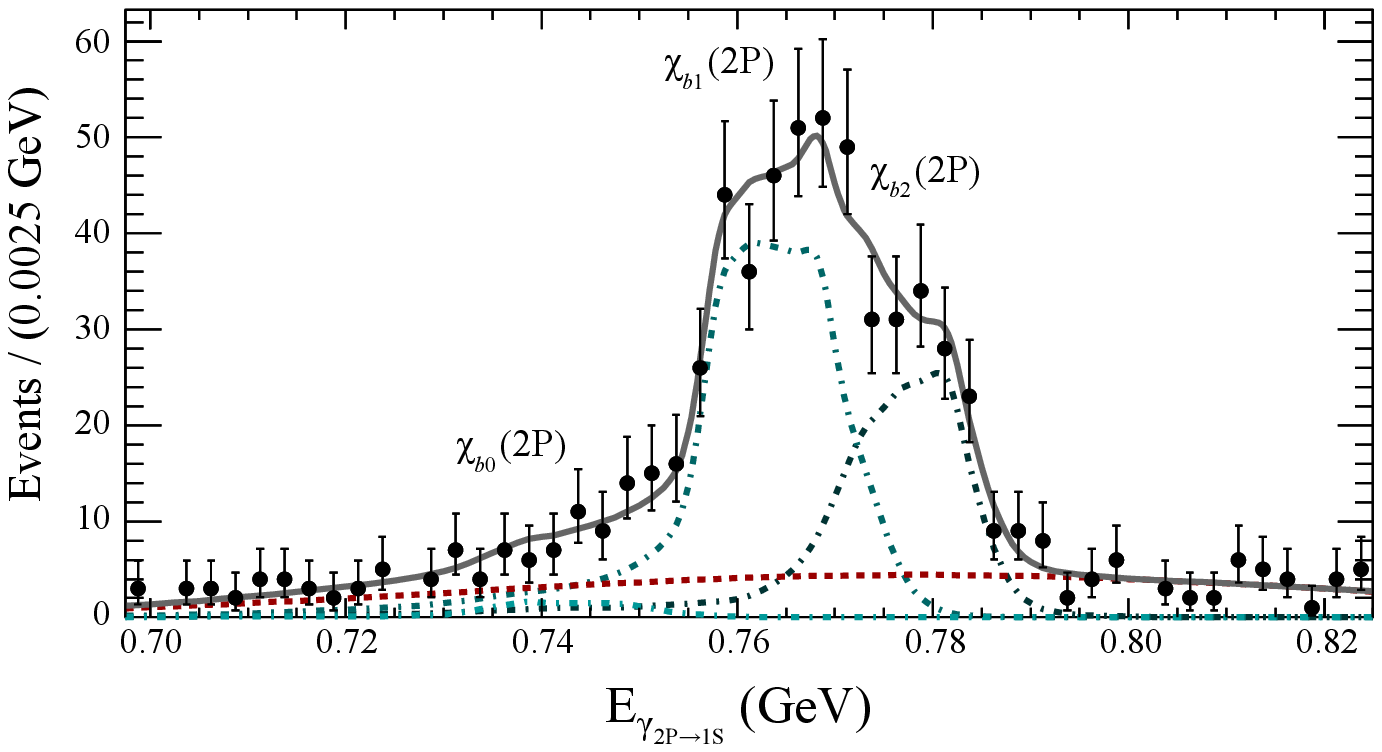}
    \caption{Fit to the photon energy $E_{2P\to 1S}$ in the $3S\to 2P\to 1S$ cascade for the conversion analysis. The data are represented by points, the total fit by a solid curve, the background component with a dotted curve, and the individual signal components of the fit by peaking dot-dashed curves in progressively darker shades for $J=0,1,2$. Clear signals for the $J=2$ and $J=1$ signals are seen.}
    \label{fig:convhi}
\end{figure*}

Figures \ref{fig:conv2s}, \ref{fig:conv3s}, \ref{fig:convlo}, and \ref{fig:convhi} show the results of the fits to the data. Compared to the calorimeter-based analysis, the statistical uncertainty in the converted-photon analysis is large and the systematic uncertainties do not readily cancel. Therefore, we quote the full product of branching fractions without normalization. The following section outlines the systematic uncertainties associated with these measurements.

\subsection{Systematic Uncertainties}
\label{sec:conv_systUnc}
The uncertainty on the luminosity is taken from the standard {\babar\ }determination \cite{ref:lumi}, which amounts to $0.58\%$ $(0.68\%)$ for $\Upsilon(3S)$ $(\Upsilon(2S))$.
The derivation of branching fractions relies on efficiencies derived from MC simulation. There are several corrections ({\em e.g.} related to particle identification, reconstruction efficiency, etc.), with accompanying uncertainties, necessary to bring simulation and data into agreement. These are determined separately from this analysis, and are employed generally by all {\babar\ }analyses. For muon identification, decay chain-dependent correction factors were estimated for each measurement, and found to be no larger than $1.3\%$, with fractional uncertainties of up to $3.3\%$. An efficiency uncertainty of $1.8\%$ is used for the calorimeter photon, and $3.3\%$ for the converted photon with a correction of $3.8\%$ (as determined in Ref.~\cite{ref:BABAR_conv}).
Uncertainty due to applying the energy scale shift and resolution smearing to the $3S\to 1P\to 1S$ cascades is estimated by varying the shift and smearing over the full range of values measured by the other decay modes.
The largest deviations from the nominal fit yields are taken as the uncertainty.
For the $J=2 (1)$ signal, the values are $^{+2.0}_{-1.7}\%$ $(^{+8.7}_{-11.5}\%)$. 

The largest source of systematic uncertainty comes from the line shape used in the fit. To account for the possibility that the MC simulation does not represent the data, the data are refit using double-sided Crystal Ball functions for the signal line shapes for all but the low-statistics $3S\to 1P\to 1S$ mode. All of the Crystal Ball shape parameters are allowed to float in the fit to the data except for the exponential power in the tail, to which the final fit result is found to be insensitive. The parameters of the background Gaussian function are allowed to float as well. The difference between the nominal fit and alternative fit yields is taken as a systematic uncertainty. These values range from $2\%$ to as large as $50\%$ for the low-yield $J=0$ channels. The yield-weighted average of all these differences is calculated using the non-$3S\to 1P\to 1S$ decay modes, and then conservatively symmetrized and applied to the $3S\to 1P\to 1S$ case. This value is found to be $5.9\%$.

To test the robustness of the low-statistics $3S\to 1P\to 1S$ fit procedure, many thousands of fits were performed on generated datasets. The output results of these tests, with signal and background yields and shapes varied, were compared with the input values. These tests determined a small background-yield-dependent bias at the level of $0.96$ events ($2.1\%$) for $J=2$ and $1.8$ events ($17\%$) for $J=1$. These values are added to the yields from the fit to the data. This correction is much smaller than the statistical uncertainty.

Finally, converting the full product of branching fractions into its constituents accrues systematic uncertainties from the measured values of the daughter branching fractions. These are taken from the PDG \cite{ref:PDG}, and can be as large as $20\%$ for some radiative bottomonium transitions. The particular values used will be stated explicitly in the text in the following sections as necessary.

\section{\boldmath Results}
\label{sec:results}
Results from this analysis are grouped into three categories: quantities directly measured (primary results), useful quantities calculated solely from the primary results (secondary results), and combinations of the primary results and relevant results from other analyses which highlight the usefulness of the high-precision compound primary results (derived results). 

{
  \begin{table*}
    \caption{Primary results from the calorimeter- and conversions-based analyses. See Fig.~\ref{fig:Bottomonium} for an explanation of the cascade and state notation. The first quantity is the branching-ratio product as defined in Eq.~\ref{eqn:brratio} from the calorimeter-based analysis. The second quantity is the fine mass splitting $\Delta M^{nP}_{J-J^{\prime}} = M^{nP}_J - M^{nP}_{J^{\prime}}$ from the calorimeter-based analysis. The third quantity is the absolute product branching fraction $\BR$ including the $\Upsilon(ns)\to\mumu$ decay from the conversions-based analysis.}
    \begin{center}
    \renewcommand{\tabcolsep}{6pt}
    \renewcommand{\arraystretch}{1.4}
	\normalsize
        \begin{tabular}{ l l l l }
          Quantity                                                             & Cascade or state & $J$        &  Value  \\ \hline \hline \\ [-2.0ex]
	$\mathscr{F}^{J/1}_{\textrm{cascade}}$  & $2S\to1P\to1S$   & $0$   &  $(3.28\pm0.37)\%$  \\
	                                                                               &                                   & $2$   &  $(55.6\pm1.6)\%$     \\
	                                                                               & $3S\to2P\to2S$  & $0$   &  $(2.31\pm0.56)\%$   \\
	                                                                               &                                   & $2$    &  $(46.9\pm2.0)\%$       \\
									   & $3S\to2P\to1S$  & $0$   & $(1.71\pm0.80)\%$     \\
                                                                                        &                                   & $2$   & $(66.6\pm3.0)\%$       \\ \hline
	$\Delta M^{\textrm{state}}_{1-J}$            & $1P$                        & $0$  & $32.49\pm0.93\mevcc$ \\
	 								   &                                   & $2$  & $-19.01\pm0.24\mevcc$ \\
									   & $2P$                        & $0$  & $23.8\pm1.7\mevcc$       \\
									   &                                     & $2$    & $-13.04\pm0.26\mevcc$ \\	\hline \hline \\ [-2.0ex]
	$\BR(\textrm{cascade})$                              & $2S\to1P(J)\to1S(\to\mumu)$ & $0$  & $(0.29^{+0.17+0.01}_{-0.14-0.08})\times 10^{-4}$ \\
									   &                                      & $1$  & $(6.86^{+0.47+0.44}_{-0.45-0.35})\times 10^{-4}$ \\
									   &                                      & $2$  & $(3.63^{+0.36+0.18}_{-0.34-0.19})\times 10^{-4}$ \\
									   & $3S\to2P(J)\to2S(\to\mumu)$ & $0$ & $(0.66^{+0.49+0.20}_{-0.40-0.03})\times 10^{-4}$ \\
									   &                                      & $1$  & $(4.95^{+0.75+1.01}_{-0.70-0.24})\times 10^{-4}$ \\
									   &                                      & $2$  & $(3.22^{+0.58+0.16}_{-0.53-0.71})\times 10^{-4}$ \\
									   & $3S\to2P(J)\to1S(\to\mumu)$ & $0$ & $(0.17^{+0.15+0.01}_{-0.14-0.12})\times 10^{-4}$ \\
									   &                                      & $1$  & $(3.52^{+0.28+0.17}_{-0.27-0.18})\times 10^{-4}$ \\
									   &                                      & $2$  & $(1.95^{+0.22+0.10}_{-0.21-0.16})\times 10^{-4}$ \\
						                               & $3S\to1P(J)\to1S(\to\mumu)$ & $1$  & $(1.16^{+0.78+0.14}_{-0.67-0.16})\times 10^{-5}$ \\
						                               &                                       & $2$  & $(4.68^{+0.99}_{-0.92}\pm0.37)\times 10^{-5}$ \\

        \end{tabular}
      \label{tab:primResults}
  \end{center}
\end{table*}
}

\subsection{Primary results} 
\label{sec:primaryResults}
We present all primary results from the calorimeter- and conversions-based analyses in Table~\ref{tab:primResults}. These include three measured quantities. First, the branching-ratio product $\mathscr{F}$, defined in Eq.~(\ref{eqn:brratio}), from the calorimeter-based analysis. Second, mass splittings $\Delta M^{nP}_{J-J^{\prime}} = M^{nP}_J - M^{nP}_{J^{\prime}}$ from the calorimeter-based analysis, where in the $2P$ case we have combined the splittings from the two $3S$ cascades to obtain an inverse-variance-weighted mean.  Third, from the conversions-based analysis we include absolute branching-fraction products $\BR$ (with statistical and systematic uncertainties), including the decay of $\Upsilon\rightarrow\mu^{+}\mu^{-}$ .

\subsection{Secondary results: Fine structure parameters for potential models} \label{sec:splittingPotential}
The spin-dependent forces in quarkonium interactions arise as spin-orbit and tensor terms in the quark-antiquark potential~\cite{ref:slac_splittings}:
\begin{equation}
  V_{SD} = a\mathbf{L\cdot S} + bS_{12},
\end{equation}
where $S_{12}=6(\mathbf{S_1\cdot \hat{r}})(\mathbf{S_2\cdot \hat{r}})-2\mathbf{S_1\cdot S_2}$. The coefficients $a$ and $b$ are interpreted as arising from scalar and vector fields, corresponding to the long-range confinement (linear) and short-range gluon-exchange (Coulomb-like) terms in the potential. 

The masses of the triplet $P$ states are split from the spin-weighted center of mass $\overline{M_{nP}} = \frac{1}{9}(M_{nP(0)}+3M_{nP(1)}+5M_{nP(2)})$ according to:
\begin{equation}
  M_{nP(J)} = \overline{M_{nP}} + a\left < \mathbf{L \cdot S} \right > + b \left < S_{12} \right >.
\end{equation}
For the triplet $P$-wave states $\{ \chi_{b0},\chi_{b1},\chi_{b2}\}$ the expectation values are $\left < \mathbf{L} \cdot \mathbf{S} \right > = \{ -2, -1, 1 \}$, $\left < S_{12} \right > = \{ -4,2, -2/5 \}$~\cite{ref:davies}. The spin-spin term $\left < \mathbf{S_1} \cdot \mathbf{S_2} \right > = \{ 1/4, 1/4, 1/4 \}$ does not differentiate between the $\chi_{bJ}$ states and therefore is ignored. 

We cast these parameters in terms of mass splittings $\Delta M_{1-0}$ and $\Delta M_{2-1}$ using the above matrix elements, thus bypassing the systematic uncertainties associated with absolute masses which are typically used:
\begin{equation}
  a=\frac{1}{6} \left ( \Delta M_{1-0}+\frac{5}{2}\Delta M_{2-1} \right ), 
\end{equation}\begin{equation}
  b=\frac{5}{72} \left ( 2\Delta M_{1-0} - \Delta M_{2-1} \right ). 
\end{equation}
This formulation provides an advantage in constraining the fine splitting parameters in potential models. Frequently the ratio
\begin{equation}\label{eqn:Rx}
  R_{\chi} = \frac{2a-\frac{12}{5}b}{a+6b} = \frac{\Delta M_{2-1}}{\Delta M_{1-0}}
\end{equation}
has been used as a convenient parameter to probe theoretical predictions in a way that is sensitive to the underlying models~\cite{ref:skwarnicki}. This allows a cancellation of experimental mass determination systematic uncertainties but is not sufficient to determine $a$ and $b$ independently.

Previous measurements of these ratios have not been in universal agreement. A persistent question is whether $R_{\chi(2P)}>R_{\chi(1P)}$, which is contrary to most models but consistent with many of the experimental determinations (see Table~\ref{tab:splitParams}). We report measurements of the parameters $a$, $b$ and $R_{\chi}$ in Table~\ref{tab:splitParams} using the primary results from the EMC-based analysis, and we find weak evidence ($0.8\sigma$) that $R_{\chi(1P)}$ is larger than $R_{\chi(2P)}$. 

{
  \begin{table*}
    \caption{Comparison of fine splitting parameters in the $1P$ and $2P$ systems. Results are compared to Besson and Skwarnicki~\cite{ref:skwarnicki} in column 4 and two more-recent measurements where these parameters were explicitly reported. See Besson and Skwarnicki~\cite{ref:skwarnicki} for a comparison of a large number of theoretical predictions.}
    \begin{center}
    \renewcommand{\tabcolsep}{4pt}
    \renewcommand{\arraystretch}{1.2}
        \begin{tabular}{ l l D{,}{\,\pm\,}{-1} D{,}{\,\pm\,}{-1} D{,}{\,\pm\,}{-1} D{,}{\,\pm\,}{-4.4} }
          Parameter  
          & $nP$ 
          & \multicolumn{1}{c}{This analysis}   
          & \multicolumn{1}{c}{Ref.~\cite{ref:skwarnicki}}
          & \multicolumn{1}{c}{CLEO 2005~\cite{ref:CLEO_3S_2S_inclusive_soft}}
          & \multicolumn{1}{c}{CLEO 1999~\cite{ref:CLEO_2S_inclusive_splittings}}  \\ \hline \hline \\ [-2.0ex]
          $a(\mev)$  & $1P$ & 13.34,0.18  & 14.2,0.8 &                 &               \\
                       & $2P$ & 9.40,0.31   & 9.4,0.2  &                 &               \\ \hline
          $b(\mev)$  & $1P$ & 3.19,0.13   & 3.0,0.3  &                 &               \\
                       & $2P$ & 2.39,0.25   & 2.3,0.1  &                 &               \\ \hline
          $R_{\chi}$ & $1P$ & 0.585,0.018 & 0.65,0.03& 0.574,0.012 & 0.54,0.03 \\  
                       & $2P$ & 0.549,0.042 & 0.58,0.01& 0.584,0.014 &               \\  
        \end{tabular}
      \label{tab:splitParams}
  \end{center}
\end{table*}
}
\subsection{Secondary results: Spin-dependent coefficients for lattice NRQCD} 
\label{sec:splittingNRQCD}
Analogous to the discussion in the previous section, the leading order spin-dependent terms in the NRQCD Hamiltonian used in lattice calculations are parametrized by the coefficients $c_3$ and $c_4$ (which are simply related to $a$ and $b$). The argument in the lattice literature has utilized cancellation of the spin-orbit and tensor expectation values shown previously to isolate $c_3$ in the $\chi_{bJ}(nP)$ mass combination
\begin{equation}
  -2M_{nP(0)} - 3M_{nP(1)} + 5M_{nP(2)}
\end{equation}
with no $c_4$ contribution. Similarly, the combination
\begin{equation}
   2M_{nP(0)} - 3M_{nP(1)} + M_{nP(2)}
\end{equation}
is proportional to the tensor coefficient $c_4^2$ and independent of $c_3$~\cite{ref:gray, ref:monahan, ref:davies}. The $P$-wave splittings in bottomonium therefore can provide a very clean check of the dominant spin-dependent terms which are useful in a wide array of NRQCD problems. We recast the above mass combinations in terms of the mass splittings $\Delta M_{1-0}$ and $\Delta M_{2-1}$ to avoid the systematic uncertainty associated with calculating the absolute mass. For the spin-orbit term we obtain
\begin{equation}
  -2\Delta M_{1-0}-5\Delta M_{2-1} = 12a 
\end{equation}
for which we measure values of $(160.0\pm2.2)\mevcc$ and $(112.7\pm3.8)\mevcc$ for the $1P$ and $2P$ triplets, respectively. The tensor term becomes
\begin{equation}
  -2\Delta M_{1-0}+\Delta M_{2-1} = -\frac{72}{5}b 
\end{equation}
for which we measure $(-46.0\pm1.9)\mevcc$ and $(-34.5\pm3.5)\mevcc$ for $1P$ and $2P$, respectively. 

\subsection{Derived results: Branching fractions}
\label{sec:BFs}
Tables~\ref{tab:Y2S_tension}, \ref{tab:Y3S_2S_tension}, and \ref{tab:Y3S_1S_tension} show the derived branching fractions using results from the $2S\to 1P\to 1S$, $3S\to 2P \to 2S$ and $3S\to 2P \to 1S$ calorimeter- and conversion-based analyses. These derived results are provided as a service in order to present the most up-to-date branching fractions using the highest quality results from multiple external sources in combination with the primary results from this analysis. Only the primary and secondary results in the previous sections should be cited as direct measurements from this analysis. We provide details in this section to allow reproduction of our derived values. 

The base measurements with the lowest uncertainty are $\mathscr{F}$ for the calorimeter-based analysis and $\mathcal{B}(mS\to nP)\cdot\mathcal{B}(nP\to pS)\cdot\mathcal{B}(pS\to\mumu)$ for the conversion-based analysis. The data used to obtain the primary results from the two analyses are complementary and share no systematic uncertainties; therefore we combine results with a standard inverse-variance-weighted mean. To derive the remaining results from these measured values, the world averages for $\mathcal{B}(mS\to nP)$ and $\mathcal{B}(nP\to pS)$ were recalculated using the method prescribed by the PDG \cite{ref:PDG}. In doing so, the measured values from experiments that measured the full exclusive decay chain were taken from the original source material and rescaled using the most up-to-date $\mathcal{B}(\Upsilon(pS)\to \ell^{+}\ell^{-})$ and $\mathcal{B}(mS\to nP)$ daughter branching fractions \cite{ref:PDG}. For $3S\to 2P\to 2S$ and $3S\to 2P\to 1S$ we use Refs.~\cite{ref:CUSB_3S_exclusive2, ref:BABAR_conv, ref:CLEO_3S_exclusive, ref:CLEOIII_exclusive3S}, and for $2S\to 1P\to 1S$ we use Refs.~\cite{ref:CLEO_3S_2S_exclusive, ref:BABAR_conv, ref:CUSB_2S_exclusive, ref:CrystalBall_2S_exclusive}. In the case of the $3S\to 1P\to 1S$ transitions, shown in Table \ref{tab:Y3S_1P_tension}, the best $\BR(1P \to 1S)$ values are calculated by including the results from the other measurements in this experiment in Table \ref{tab:Y2S_tension} in the overall average to find $\BR(1P(2) \to 1S)=(18.7\pm1.1)\%$ and $\BR(1P(1) \to 1S)=(35.0\pm2.3)\%$. Systematic and statistical uncertainties are summed in quadrature; in cases where the positive and negative uncertainties differ we select the higher of the two with negligible effect. 

The strategy employed in the calorimeter-based analysis aims specifically at reaching the $1P(0)\to 1S$, $2P(0)\to 2S$ and $2P(0)\to 1S$ transitions, which we have seen with significances of $8.9$, $5.9$ and $2.1\sigma$. The primary result of the conversion-based analysis is the measurement of $3S\to 1P\to 1S$. The results are in good agreement with both the previous CLEO \cite{ref:CLEO_3S_2S_exclusive} and {\babar\ }\cite{ref:BABAR_conv} results, with central values falling between those measurements (albeit with larger overall uncertainties). All other derived branching fractions are largely consistent with previous results.

{
  \begin{table*}[p]
    \caption{Summary of primary and derived results from the $2S\to1P\to1S$ analysis. Columns 2 and 3 contain the results from the calorimeter-based and converted photon analyses, with combined results in column 4. The significance of the derived quantities for this analysis is shown in column 5. For comparison purposes, the final column gives the global average determined from our rescaling and combination of previous results, as described in the text. }
    \begin{center}
    \renewcommand{\arraystretch}{1.2}
    \renewcommand{\tabcolsep}{6pt}
      \begin{tabular}{ l D{,}{\,\pm\,}{-1} D{,}{\,\pm\,}{-1} D{,}{\,\pm\,}{-1} c D{,}{\,\pm\,}{-4.4}}
        Measured Quantity                     
        & \multicolumn{3}{c}{\babar\ Measurement $(10^{-2})$}                
        & $(\sigma)$
        & \multicolumn{1}{c}{Previous} \\                                     
        & \multicolumn{1}{c}{EMC}            
        & \multicolumn{1}{c}{Converted}               
        & \multicolumn{1}{c}{Combined}        
        &       
        & \multicolumn{1}{c}{Average $(10^{-2})$} \\ \hline \hline \\ [-2.0ex]
        $\BR(2S\to1P(0))$                     & 4.42,1.04          & 6.8,4.1  & 4.6,1.0     & 4.5 & 3.8,0.4    \\
        $\BR(2S\to1P(1))$                     & 7.20,0.65          & 8.16,0.97  & 7.50,0.54     & 14  & 6.9,0.4    \\
        $\BR(2S\to1P(2))$                     & 6.85,0.58          & 7.7,1.0  & 7.06,0.51     & 14  & 7.15,0.35  \\ \hline
        $\BR(1P(0)\to1S)$                     & 2.02,0.32          & 3.1,1.8  & 2.06,0.32     & 6.5 & 1.73,0.35  \\
        $\BR(1P(1)\to1S)$                     & 35.4,2.8           & 40.1,4.5  & 36.7,2.4      & 15  & 33.9,2.4   \\   
        $\BR(1P(2)\to1S)$                     & 18.2,1.3           & 20.5,2.5   & 18.7,1.1      & 16  & 19.0,1.3   \\ \hline
        $\BR(2S\to1P(0))/\BR(2S\to1P(1))$     & 64.1,15.5          & 84,50  & 66,15     & 4.4 & 55.1,6.6            \\
        $\BR(2S\to1P(2))/\BR(2S\to1P(1))$     & 99.3,10.2          & 94,15  & 97.8,8.5      & 12  & 103.6,7.9             \\
        $\BR(1P(0)\to1S)/\BR(1P(1)\to1S)$     & 5.96,0.98          & 7.8,4.5  & 6.04,0.96     & 6.3 & 5.12,1.09             \\ 
        $\BR(1P(2)\to1S)/\BR(1P(1)\to1S)$     & 53.7,4.4           & 51.1,7.6   & 53.0,3.8      & 14  & 56.0,5.5            \\ 
        $\BR(2S\to1P(0))\cdot\BR(1P(0)\to1S)$ & 0.0767,0.0092      & 0.118,0.066& 0.0775,0.0091 & 8.5 & 0.066,0.011        \\ 
        $\BR(2S\to1P(1))\cdot\BR(1P(1)\to1S)$ & 2.44,0.14          & 2.77,0.26  & 2.51,0.12     & 21  & 2.339,0.097          \\ 
        $\BR(2S\to1P(2))\cdot\BR(1P(2)\to1S)$ & 1.30,0.07          & 1.46,0.17  & 1.32,0.061    & 22  & 1.357,0.064        \\ \hline
        $\mathscr{F}^{0/1}(2S\to1P\to1S)$     & 3.28,0.37 & 4.3,2.4                     & 3.30,0.37             & 9.0 & 2.89,0.52           \\
        $\mathscr{F}^{2/1}(2S\to1P\to1S)$     & 55.6,1.6  & 52.9,7.9                     & 55.5,1.6             & 35& 58.3,4.7            \\ 
      \end{tabular}
      \label{tab:Y2S_tension}
    \end{center}
  \end{table*}
}

{
  \begin{table*}[p]
    \caption{Summary of primary and derived results from the $3S\to2P\to2S$ analysis. See the caption of Table~\ref{tab:Y2S_tension} for details.}
    \begin{center}
    \renewcommand{\arraystretch}{1.2}
    \renewcommand{\tabcolsep}{6pt}
     \begin{tabular}{ l D{,}{\,\pm\,}{-1} D{,}{\,\pm\,}{-1} D{,}{\,\pm\,}{-1} c D{,}{\,\pm\,}{-4.4}}
        Measured Quantity                     
        & \multicolumn{3}{c}{\babar\ Measurement $(10^{-2})$}                
        & $(\sigma)$
        & \multicolumn{1}{c}{Previous} \\                                     
        & \multicolumn{1}{c}{EMC}            
        & \multicolumn{1}{c}{Converted}               
        & \multicolumn{1}{c}{Combined}        
        &       
        & \multicolumn{1}{c}{Average $(10^{-2})$} \\ \hline \hline \\ [-2.0ex]
        $\BR(3S\to2P(0))$                     & 4.5,1.7          & 20,17  & 4.6,1.7   & 2.7 & 5.9,0.6   \\
        $\BR(3S\to2P(1))$                     & 14.7,2.6           & 14.1,4.4   & 14.5,2.2     & 6.6 & 12.6,1.2    \\
        $\BR(3S\to2P(2))$                     & 11.2,2.2           & 17.4,5.6   & 12.0,2.0    & 6.0 & 13.1,1.6    \\ \hline
        $\BR(2P(0)\to2S)$                     & 1.29,0.30          & 5.8,4.7  & 1.31,0.30   & 4.4 & 1.71,0.56  \\
        $\BR(2P(1)\to2S)$                     & 21.2,2.8           & 20.3,5.8  & 21.1,2.5    & 8.5 & 18.2,2.8   \\   
        $\BR(2P(2)\to2S)$                     & 8.2,1.8          & 12.7,4.0   & 9.0,1.7    & 5.4 & 9.6,1.4   \\ \hline
        $\BR(3S\to2P(0))/\BR(3S\to2P(1))$     & 35,14          & 143,130    & 37,14     & 2.6 & 46.8,6.5             \\
        $\BR(3S\to2P(2))/\BR(3S\to2P(1))$     & 89,19          & 124,46     & 94,18     & 5.3 & 104,16               \\
        $\BR(2P(0)\to2S)/\BR(2P(1)\to2S)$     & 7.1,1.5          & 29,24  & 7.2,1.5     & 4.6 & 9.4,3.4    \\ 
        $\BR(2P(2)\to2S)/\BR(2P(1)\to2S)$     & 45.1,7.2           & 63,22  & 46.9,6.9      & 6.8 & 53,11   \\ 
        $\BR(3S\to2P(0))\cdot\BR(2P(0)\to2S)$ & 0.076,0.016      & 0.34,0.27& 0.077,0.016 & 4.9 & 0.101,0.032\\ 
        $\BR(3S\to2P(1))\cdot\BR(2P(1)\to2S)$ & 2.68,0.24          & 2.56,0.69  & 2.66,0.22     & 12  & 2.30,0.27  \\ 
        $\BR(3S\to2P(2))\cdot\BR(2P(2)\to2S)$ & 1.08,0.14          & 1.67,0.48  & 1.12,0.13     & 8.6 & 1.255,0.097  \\ \hline
        $\mathscr{F}^{0/1}(3S\to2P\to2S)$     & 3.31,0.56 & 13,11          & 3.33,0.56         &  6.0& 4.4,1.5             \\
        $\mathscr{F}^{2/1}(3S\to2P\to2S)$     & 46.9,2.0  & 65,26          &   47.0,2.0          &  24& 54.7,7.7           \\ 
      \end{tabular}
      \label{tab:Y3S_2S_tension}
  \end{center}
\end{table*}
}

{
  \begin{table*}[p]
\caption{Summary of primary and derived results from the $3S\to2P\to1S$ analysis. See the caption of Table~\ref{tab:Y2S_tension} for details.}
    \begin{center}
    \renewcommand{\arraystretch}{1.2}
    \renewcommand{\tabcolsep}{6pt}
   \begin{tabular}{ l D{,}{\,\pm\,}{-1} D{,}{\,\pm\,}{-1} D{,}{\,\pm\,}{-1} c D{,}{\,\pm\,}{-4.4}}
        Measured Quantity                     
        & \multicolumn{3}{c}{\babar\ Measurement $(10^{-2})$}                
        & $(\sigma)$
        & \multicolumn{1}{c}{Previous} \\                                     
        & \multicolumn{1}{c}{EMC}            
        & \multicolumn{1}{c}{Converted}               
        & \multicolumn{1}{c}{Combined}        
        &       
        & \multicolumn{1}{c}{Average $(10^{-2})$} \\ \hline \hline \\ [-2.0ex]
        $\BR(3S\to2P(0))$                     & 2.0,1.7    & 6.5,8.5  & 2.2,1.6     & 1.3  & 5.9,0.6      \\
        $\BR(3S\to2P(1))$                     & 13.7,1.7     & 14.8,2.1   & 14.1,1.4    & 10.5  & 12.6,1.2       \\
        $\BR(3S\to2P(2))$                     & 12.0,1.7     & 11.7,2.2   & 11.9,1.4    & 8.7  & 13.1,1.6     \\ \hline
        $\BR(2P(0)\to1S)$                     & 0.35,0.17  & 1.1,1.3  & 0.37,0.17 & 2.2  & 1.04,0.72   \\
        $\BR(2P(1)\to1S)$                     & 10.5,1.2    & 11.3,1.5   & 10.78,0.94    & 11.4  & 9.6,1.1      \\   
        $\BR(2P(2)\to1S)$                     & 6.16,0.88    & 6.0,1.1  & 6.10,0.69   & 8.9  & 6.71,0.87    \\ \hline
        $\BR(3S\to2P(0))/\BR(3S\to2P(1))$     & 16,13    & 44,57  & 17,13   & 1.3  & 46.8,6.5                \\
        $\BR(3S\to2P(2))/\BR(3S\to2P(1))$     & 95,17    & 79,18  & 88,12   & 7.2  & 104,16                        \\
        $\BR(2P(0)\to1S)/\BR(2P(1)\to1S)$     & 3.7,1.8    & 10,11  & 3.8,1.8   & 2.2  & 10.8,7.6   \\ 
        $\BR(2P(2)\to1S)/\BR(2P(1)\to1S)$     & 64,10    & 53,11  & 59.2,7.7    & 7.7  & 70,12      \\ 
        $\BR(3S\to2P(0))\cdot\BR(2P(0)\to1S)$ & 0.0207,0.0097& 0.067,0.075 &0.021,0.010 & 2.2  & 0.061,0.042 \\ 
        $\BR(3S\to2P(1))\cdot\BR(2P(1)\to1S)$ & 1.320,0.085    & 1.42,0.14  &1.348,0.072    & 19 & 1.21,0.049           \\ 
        $\BR(3S\to2P(2))\cdot\BR(2P(2)\to1S)$ & 0.807,0.057  & 0.79,0.11&0.803,0.050  & 16 & 0.879,0.040   \\ \hline
        $\mathscr{F}^{0/1}(3S\to2P\to1S)$     & 1.71,0.80 & 4.7,5.3          & 1.78,0.79             & 2.2  & 5.1,3.5                        \\
        $\mathscr{F}^{2/1}(3S\to2P\to1S)$     & 66.6,3.0  & 55.4,9.3          & 65.5,2.9             & 23 & 72.6,4.4                       \\ 
      \end{tabular}
      \label{tab:Y3S_1S_tension}
  \end{center}
\end{table*}
}

{
  \begin{table*}[p]
\caption{Summary of the results from the $3S\to1P\to1S$ analysis in comparison to other measurements. Values for $\BR(3S\to1P)$ have been rescaled using our best averages for $\BR(1P\to 1S)$.}
    \begin{center}
    \renewcommand{\arraystretch}{1.2}
    \renewcommand{\tabcolsep}{8pt}
      \begin{tabular}{ l l l l }
        Measured Quantity                     & This work  & CLEO~\cite{ref:CLEO_3S_2S_exclusive} & {\babar\ }2011~\cite{ref:BABAR_conv} \\ \hline \hline \\ [-2.0ex]
                                              & $(10^{-5})$      &  & \\
        $\BR(3S\to1P(1)\to1S(\to\mumu))$      & $1.16^{+0.79}_{-0.68}$ & $1.33\pm0.38$ & $$ \\
        $\BR(3S\to1P(2)\to1S(\to\mumu))$      & $4.68^{+1.06}_{-0.99}$ & $3.56\pm0.57$ & $$     \\ \hline
                                              & $(10^{-3})$      & & \\
        $\BR(3S\to1P(1))$                     & $1.3^{+0.9}_{-0.8}$    & $1.60\pm0.47$ & $0.5^{+0.4}_{-0.3}$ \\
        $\BR(3S\to1P(2))$                     & $10.1^{+2.5}_{-2.3}$    & $7.76\pm1.35$ & $10.5^{+0.8}_{-0.7}$ \\
      \end{tabular}
      \label{tab:Y3S_1P_tension}
  \end{center}
\end{table*}
}

\subsection{Derived results: electric dipole matrix elements} 
\label{sec:matrixElements}
{
  \begin{table*}
    \caption{Photon energies in the parent rest frame for transitions measured in this analysis, calculated by correcting the spin-weighted average of Gaussian means $\mu_J$ to the correct average as calculated from PDG masses~\cite{ref:PDG}. Only $J=1$ values and mass splittings are used for the matrix element ratio calculations according to Eq.~(\ref{eqn:widthRatio}).} 
    \begin{center}
    \renewcommand{\tabcolsep}{8pt}
    \renewcommand{\arraystretch}{1.2}
        \begin{tabular}{ l D{,}{\,\pm\,}{-4.4} }
          Transition  & \multicolumn{1}{c}{$E_\gamma(\mev)$} \\ \hline \hline \\ [-2.0ex]
          2S\to 1P(0) & 162.8,1.0    \\
          2S\to 1P(1) & 130.34,0.45  \\
          2S\to 1P(2) & 111.33,0.49  \\
          3S\to 2P(0) & 123.4,1.8    \\ 
          3S\to 2P(1) & 99.61,0.64   \\
          3S\to 2P(2) & 86.59,0.66   \\
          1P(0)\to 1S & 400.1,1.0    \\
          1P(1)\to 1S & 432.62,0.45  \\
          1P(2)\to 1S & 451.63,0.49  \\
          2P(0)\to 2S & 208.6,1.8    \\
          2P(1)\to 2S & 232.44,0.51  \\
          2P(2)\to 2S & 245.35,0.53  \\
          2P(0)\to 1S & 771.5,1.8    \\
          2P(1)\to 1S & 795.29,0.48  \\
          2P(2)\to 1S & 808.31,0.50  \\
        \end{tabular}
      \label{tab:derivedEnergies}
  \end{center}
\end{table*}
}

In the nonrelativistic limit the $E1$ matrix elements of Eq.~(\ref{eqn:gammaE1}) depend only on the primary and orbital quantum numbers $n$ and $L$; ratios of matrix elements for two transitions that differ only in spin are thus convenient probes of relativistic corrections. We use the derived branching ratios from the previous section to calculate matrix element ratios with the $J=1$ element in the denominator:
\begin{equation}
  \label{eqn:widthRatio}
  \begin{split}
    \frac{| \langle f_J | r | i_J \rangle |^2}{| \langle f_1 | r | i_1 \rangle |^2} &= \frac{3}{2J_{f}+1} \frac{E_1^3}{E_J^3} \frac{\BR(i_J\to f_J)}{\BR(i_1\to f_1)} \\
    &=\frac{3}{2J+1}\left ( 1 - \frac{\Delta M_{J-1}}{E_1} \right )^{-3} \frac{\BR(i_J\to f_J)}{\BR(i_1\to f_1)} 
  \end{split}
\end{equation}
where in the last line we have recast the absolute photon energies in reference to the $J=1$ energy and the measured mass splittings to minimize uncertainties; $\Delta M_{J-1}$ is $-\Delta M_{1-0}$ or $\Delta M_{2-1}$, depending on the ratio considered. 

The calculation of the matrix element ratios requires branching ratios which we do not explicitly measure. Instead we use the best available branching fractions as shown in Tables~\ref{tab:Y2S_tension},~\ref{tab:Y3S_2S_tension},~\ref{tab:Y3S_1S_tension}. Similarly for the absolute line energies we calculate the spin-weighted center of energy for the $1P$ and $2P$ states $\overline{\mu^{mP}}$ and the spin-weighted center of mass $\overline{M^{mP}}$ from PDG~\cite{ref:PDG} masses similarly. We derive all line energies using $E_J = \overline{M^{mP}} \pm (\overline{\mu^{mP}} - \mu_J)$ where the sign depends on the transition. The derived line energies are shown in Table~\ref{tab:derivedEnergies}, although only the $J=1$ values are used in the matrix element ratio calculations, shown in Table~\ref{tab:widthRatios}.

{
  \begin{table*}
    \caption{Matrix element ratios derived using combined results from this analysis from Tables~\ref{tab:Y2S_tension}, \ref{tab:Y3S_2S_tension} and \ref{tab:Y3S_1S_tension}. Column four contains the world averages as of 1993 from Besson and Skwarnicki~\cite{ref:skwarnicki} while column five contains CLEO~\cite{ref:CLEO_2S_inclusive_splittings} and CLEO-III ~\cite{ref:CLEOIII_exclusive3S} results.}
    \begin{center}
    \renewcommand{\arraystretch}{2.0}
    \renewcommand{\tabcolsep}{6pt}
        \begin{tabular}{ c c c c c c }
          $\dfrac{n_iL_i(J_i)\to n_fL_f(J_f)}{n_i^{\prime}L_i^{\prime}(J_i^{\prime})\to n_f^{\prime}L_f^{\prime}(J_f^{\prime})}$ & $\dfrac{| \langle n_fL_f(J_f) | r | n_iL_i(J_i) \rangle |^2}{| \langle n_f^{\prime}L_f^{\prime}(J_f^{\prime}) | r | n_i^{\prime}L_i^{\prime}(J_i^{\prime}) \rangle |^2}$ & $\sigma$ from unity & Ref.~\cite{ref:skwarnicki} & Refs.~\cite{ref:CLEO_2S_inclusive_splittings} and ~\cite{ref:CLEOIII_exclusive3S} \\  \hline \hline \\ [-2.0ex]
          $\dfrac{3S\to2P(0)}{3S\to2P(1)}$                                     & $0.39\pm0.18$   & $-3.4$  & $0.74\pm0.06$ & $$            & \\  \\ [-4.0ex]
          $\dfrac{3S\to2P(2)}{3S\to2P(1)}$                                     & $0.85\pm0.13$   & $-1.2$  & $1.17\pm0.04$ & $$            & \\ [-4.0ex] \\ \hline \\ [-4.0ex]
          $\dfrac{2S\to1P(0)}{2S\to1P(1)}$                                     & $1.01\pm0.24$  & $0$  & $0.95\pm0.16$ & $0.75\pm0.28$ & \\ \\ [-4.0ex]
          $\dfrac{2S\to1P(2)}{2S\to1P(1)}$                                     & $0.941\pm0.089$ & $-0.7$  & $0.92\pm0.11$ & $1.02\pm0.11$ &  \\ [-4.0ex] \\ \hline \\ [-3.0ex]
          $\dfrac{2P(0)\to1S}{2P(0)\to2S} \Big/ \dfrac{2P(1)\to1S}{2P(1)\to2S}$& $0.43\pm0.23$   & $-2.5$  & $0.37\pm0.3$  & $$            & \\ \\ [-3.0ex]
          $\dfrac{2P(2)\to1S}{2P(2)\to2S} \Big/ \dfrac{2P(1)\to1S}{2P(1)\to2S}$& $1.49\pm0.39$   & $+1.3$  & $1.33\pm0.26$ & $1.21\pm0.06$ & \\ 
        \end{tabular}
      \label{tab:widthRatios}
  \end{center}
\end{table*}
}

\section{Discussion}
\label{sec:summary}
We have presented an array of primary (Table~\ref{tab:primResults}), secondary and derived spectroscopic results (Tables~\ref{tab:Y2S_tension},~\ref{tab:Y3S_2S_tension},~\ref{tab:Y3S_1S_tension},~\ref{tab:Y3S_1P_tension}) comprising a comprehensive study of electric dipole transitions between the $\Upsilon(1S,2S,3S)$ and $\chi_{bJ}(1P,2P)$ bottomonium states. These results include weak evidence for the $\chi_{b0}(2P)\to\gamma\Upsilon(1S)$ transition and the best observational significance yet for the $\chi_{b0}(2P)\to\gamma\Upsilon(2S)$ and $\chi_{b0}(1P)\to\gamma\Upsilon(1S)$ transitions, along with determinations of splitting parameter values (Table~\ref{tab:splitParams}) and a calculation of spin-dependent matrix element ratios (Table~\ref{tab:widthRatios}). 

The results for $\BR(\Upsilon(3S)\to\gamma\chi_{bJ}(1P))$ are consistent with both previous measurements from CLEO \cite{ref:CLEO_3S_2S_exclusive} and \babar\ \cite{ref:BABAR_conv}, whose level of disagreement is reduced to less than $2\sigma$ when the branching fractions are recalculated using the best overall averages with input from this analysis.
The unusual $\BR(\Upsilon(3S)\to\gamma\chi_{b2}(1P))> \BR(\Upsilon(3S)\to\gamma\chi_{b1}(1P))> \BR(\Upsilon(3S)\to\gamma\chi_{b0}(1P))$ pattern for the decay rate is seen here, in agreement with recent theoretical predictions \cite{ref:badalian}.

We see variations in the matrix element ratios at the level of one to two standard deviations, as shown in Table~\ref{tab:widthRatios} for several ratios, and a slightly greater than $3\sigma$ deviation from unity for the $(\Upsilon(3S)\to\gamma\chi_{b0}(2P))/(\Upsilon(3S)\to\gamma\chi_{b1}(2P))$ ratio. These results are competitive with previous analyses, particularly for the $2/1$ ratios.

The splitting parameter measurements in Table~\ref{tab:splitParams} are competitive and largely consistent with previous results. Our value for the splitting ratio $R_{\chi(1P)}$ supports the most-recent CLEO results~\cite{ref:CLEO_3S_2S_inclusive_soft, ref:CLEO_2S_inclusive_splittings} and is almost $2\sigma$ below the $1993$ world average~\cite{ref:skwarnicki}. Our measurements of the NRQCD parameters $c_3$ and $c_4$ are competitive with world-averages commonly used, with a distinct improvement in the determination of $c_4$. 

Inclusive analyses are better suited for the mass splitting and $\Upsilon(3S)\to\gamma\chi_{b\mathrm{J}}(2P)$ matrix element measurements and we suggest such an analysis using the relative-energy techniques we have presented, to target high-precision measurements of these quantities with explicit determinations of $a$, $b$, $c_3$ and $c_4$. 

\section{Acknowledgments}
\label{sec:acknowledgments}
We are grateful for the 
extraordinary contributions of our \pep2\ colleagues in
achieving the excellent luminosity and machine conditions
that have made this work possible.
The success of this project also relies critically on the 
expertise and dedication of the computing organizations that 
support \babar.
The collaborating institutions wish to thank 
SLAC for its support and the kind hospitality extended to them. 
This work is supported by the
US Department of Energy
and National Science Foundation, the
Natural Sciences and Engineering Research Council (Canada),
the Commissariat \`a l'Energie Atomique and
Institut National de Physique Nucl\'eaire et de Physique des Particules
(France), the
Bundesministerium f\"ur Bildung und Forschung and
Deutsche Forschungsgemeinschaft
(Germany), the
Istituto Nazionale di Fisica Nucleare (Italy),
the Foundation for Fundamental Research on Matter (The Netherlands),
the Research Council of Norway, the
Ministry of Education and Science of the Russian Federation, 
Ministerio de Econom\'{\i}a y Competitividad (Spain), the
Science and Technology Facilities Council (United Kingdom),
and the Binational Science Foundation (U.S.-Israel).
Individuals have received support from 
the Marie-Curie IEF program (European Union) and the A. P. Sloan Foundation (USA). 


\end{document}